# Flow-Based Algorithms for Local Graph Clustering[*]

Lorenzo Orecchia
orecchia@mit.edu
MIT Math

Zeyuan Allen Zhu
zeyuan@csail.mit.edu
MIT CSAIL

## Abstract

Given a subset $A$ of vertices of an undirected graph $G$, the cut-improvement problem asks us to find a subset $S$ that is similar to $A$ but has smaller conductance. An elegant algorithm for this problem has been given by Andersen and Lang [AL08] and requires solving a small number of single-commodity maximum flow computations over the whole graph $G$. In this paper, we introduce `LocalImprove`, the first cut-improvement algorithm that is *local*, i.e., that runs in time dependent on the size of the input set $A$ rather than on the size of the entire graph. Moreover, `LocalImprove` achieves this local behavior while closely matching the same theoretical guarantee as the global algorithm of Andersen and Lang.

The main application of `LocalImprove` is to the design of better local-graph-partitioning algorithms. All previously known local algorithms for graph partitioning are random-walk based and can only guarantee an output conductance of $\tilde{O}(\sqrt{\phi_{\mathsf{opt}}})$ when the target set has conductance $\phi_{\mathsf{opt}} \in [0,1]$. Very recently, Zhu, Lattanzi and Mirrokni [ZLM13] improved this to $O(\phi_{\mathsf{opt}}/\sqrt{\mathsf{Conn}})$ where the internal connectivity parameter $\mathsf{Conn} \in [0,1]$ is defined as the reciprocal of the mixing time of the random walk over the graph induced by the target set. This regime is of high practical interest in learning applications as it corresponds to the case when the target set is a well-connected ground-truth cluster. In this work, we show how to use `LocalImprove` to obtain a constant approximation $O(\phi_{\mathsf{opt}})$ as long as $\mathsf{Conn}/\phi_{\mathsf{opt}} = \Omega(1)$. This yields the first flow-based algorithm for local graph partitioning. Moreover, its performance strictly outperforms the ones based on random walks and surprisingly matches that of the best known global algorithm, which is $\mathsf{SDP}$-based, in this parameter regime [MMV12].

Finally, our results show that spectral methods are not the only viable approach to the construction of local graph partitioning algorithm and open door to the study of algorithms with even better approximation and locality guarantees.

## 1 Introduction

Graph partitioning is a fundamental algorithmic primitive with applications in numerous areas, including data mining, computer vision, social network analysis and VLSI layout. In machine learning, graph partitioning is particularly useful in the context of clustering when the data set is given by a similarity matrix, representing a graph.

Formally, given an undirected and unweighted[1] graph $G = (V, E)$, we consider the following fundamental partitioning objective, known as *conductance*:

$$\phi(S) \stackrel{\text{def}}{=} \frac{|E(S, \bar{S})|}{\min\{\operatorname{vol}(S), \operatorname{vol}(\bar{S})\}} \in [0, 1] \ ,$$

---

[*]A short version of this paper has appeared in the proceedings of SODA 2014. [OZ14]
[1]All our results can be generalized to weighted graphs.

where $\mathrm{vol}(S) \stackrel{\text{def}}{=} \sum_{v \in S} \deg(v)$. Finding $S \subset V$ with the smallest $\phi(S)$ is called the conductance minimization problem.

The problem of conductance minimization is UGC-hard to approximate within any constant factor [CKK+06]. On the positive side, spectral partitioning algorithms output a solution with conductance $O(\sqrt{\phi_{\mathsf{opt}}})$ [Alo86, SJ89]; Leighton and Rao [LR99] provide a $O(\log n)$ approximation using an LP relaxation and Arora, Rao and Vazirani [ARV09] achieve the best known $O(\sqrt{\log n})$-approximation using a powerful SDP relaxation. These results, along with recent versions with optimized running times [OSV12, OSVV08, AHK10, AK07, She09], are all obtained by *global* algorithms: their time complexity depends at least linearly on the size of $G$. Unfortunately, even such algorithms are not suitable for many applications to very large datasets. For instance, practitioners may have to deal with massive graphs containing several millions of putative clusters (e.g., webpage similarity graph and youtube video similarity graph) [GS12, GLMY11, ACE+13, AGM12].

The large size of the graph and the fact that practitioners often operate in a semi-supervised setting, in which only a certain region of the graph is investigated, are compelling motivations to develop *local* graph partitioning algorithms. This trend was initiated by Spielman and Teng [ST04], and then followed by a series of papers [ST13, ACL06, AP09, OT12, ZLM13]. These works attempt to solve the conductance minimization problem locally, with running time only dependent on the volume and the conductance of the output set $S$, and independent of any graph parameter.

All previous results on local graph partitioning are spectral in nature and are based on analyzing the behavior of random walks. To respect the locality requirement, the random walks used by these algorithms are essentially truncated after a certain number of steps. This ensures the locality of the vector of marginal probabilities, while also providing a sufficiently good approximation to the true probability distribution of the random-walk.

Despite its widespread use in global graph partitioning, flow methods did not seem to lend themselves to this kind of localization because the building blocks of a generic flow algorithm are inherently non-local, i.e. an arbitrary augmenting path from source to sink may touch almost all vertices in the graph. We overcome this obstacle by using as our primitive operation in our algorithm truncated blocking flow computations. Because blocking flows are restricted to use shortest-paths between source and sink, and not generic, potentially very long augmenting paths, this choice of basic steps enables us to keep our algorithm local. As a result, we construct the first local graph-partitioning algorithm that combines spectral and flow methods to obtain a better approximation guarantee for local graph partitioning in a regime of high practical importance, described in Section 1.2. Our main tool in developing such an algorithm is the construction of a local procedure to solve the cut-improvement problem.

## 1.1 A Flow-Based Local Algorithm for the Cut-Improvement Problem.

In the cut-improvement problem, we are given an input vertex set $A \subset V$ and are asked to find a set $S$ having conductance competitive to all other sets $S^*$ that are "well-correlated" with $A$. In other words, the cut-improvement problem can be thought of as a local-search problem, in which we seek a low-conductance cut $S$ among the cuts near, i.e. well-correlated with, cut $A$ for some appropriate notion of locality, i.e. correlation, over the space of cuts.

If one adopts a very strong definition of locality and restricts $S^*$ to be subsets of $A$, the cut-improvement problem can be solved exactly by applying the parametric-maximum-flow algorithm of Gallo, Grigoriadis, and Tarjan [GGT89]. This improvement algorithm is crucial to several theoretical results, including minimum bisection [FK02] and hierarchical oblivious routing [HHR03]. It is also applied in practice as the Max-flow Quotient-cut Improvement (MQI) algorithm [LR04] that is often used to refine the results of the METIS graph-partitioning heuristic [KK98]. It is important to notice that `MQI` is local, as it only operates on the graph induced by the input set $A$.

In a more recent paper, Andersen and Lang [AL08] weaken the definition of "well-correlated" in a natural way, allowing subsets $S^*$ to have non-zero intersection with $V - A$: for $\delta \in (0, 1]$, they say that $S^*$ $\delta$-overlaps with $A$ if $\frac{\text{vol}(S^* \cap A)}{\text{vol}(S^*)} \geq \delta$. Their `Improve` algorithm essentially outputs a set $S$ with conductance $\phi(S)$ at most a factor $O(1/\delta)$ away from $\phi(S^*)$ for all sets $S^*$ satisfying the correlation guarantee $\frac{\text{vol}(S^* \cap A)}{\text{vol}(S^*)} \geq \delta$. This guarantee holds simultaneously for all values of $\delta \in \left[\frac{\text{vol}(A)}{\text{vol}(V-A)}, 1\right]$.

In practice, this means that if there exists a low-conductance $S^*$ cut very near the input set $A$ (e.g. O(1)-overlapping with $A$), `Improve` will output a cut $S$ with $\phi(S)$ very close to $\phi(S^*)$ (e.g. a constant approximation). Similarly, if all low-conductance cuts $S^*$ have poor overlapping with $A$, the output set $S$ may have very large conductance.

The `Improve` algorithm can be easily seen as both generalizing and outperforming `MQI`, as the latter yields the same guarantee but only for cuts that 1-overlap the input set $A$. However, this improved performance comes at the cost of the loss of locality, as `Improve` requires running a small number of *global* $s - t$ flow computations over an augmented copy of the instance graph $G$ [AL08]. Notice that this is not just an issue of implementation: `Improve` must necessarily run globally as it is required to establish a guarantee for cuts $\delta$-overlapping the input set $A$ for all values of $\delta \geq \frac{\text{vol}(A)}{\text{vol}(V-A)}$, i.e. for all cuts in the graph.[2]

**Our Contribution.** We introduce a local formulation of `Improve` and provide two local algorithms, `LocalFlow` and `LocalFlow`<sup>exact</sup> that closely match the guarantee of [AL08] with different running times. To achieve the desired locality, our algorithms take as input an additional parameter $\sigma \in \left[\frac{\text{vol}(A)}{\text{vol}(V-A)}, 1\right]$ and output a cut $S$ achieving the same guarantee as `Improve`, but limited to cuts $S^*$ whose overlap with $A$ is at least $\sigma$.

**Theorem 1a** (informal). *Given set $A \subset V$ of the graph with* $\text{vol}(A) \ll \text{vol}(V)$ *and given a constant* $\sigma \in \left[\frac{\text{vol}(A)}{\text{vol}(V-A)}, 1\right]$, `LocalImprove`$_G(A, \sigma)$ *output a set $S$ such that:*

- *for any $S^* \subset V$ satisfying* $\text{vol}(S^*) \leq \text{vol}(V - S^*)$ *and $\delta$-overlapping with $A$ for $\delta \geq \sigma$,*
$$\phi(S) \leq O(1/\delta) \cdot \phi(S^*) \; ;^3$$
- $\text{vol}(S) \leq O(1/\sigma) \cdot \text{vol}(A).$[3]

*The algorithm is local, as it explores at most* $O\left(\frac{\text{vol}(A)}{\sigma}\right)$ *volume of the graph $G$. More precisely,* `LocalImprove` *runs in time* $\widetilde{O}\left(\frac{\text{vol}(A)}{\sigma \cdot \phi(S)}\right)$.

**Theorem 1b** (informal). *In the same setting as Theorem 1a,* `LocalImprove`$_G^{exact}(A, \sigma)$ *gives the same performance but runs in time* $\widetilde{O}\left(\left(\frac{\text{vol}(A)}{\sigma}\right)^{1.5}\right)$.

Effectively, our algorithms allow us to interpolate between `MQI`, when $\sigma = 1$, and `Improve`, when $\sigma = \frac{\text{vol}(A)}{\text{vol}(V-A)}$, while preserving a locality guarantee in between. We expect our algorithms to find applications both in practice, as efficient and local alternatives to `Improve`, and in theory, where they open the way to the design of local graph partitioning algorithms that achieve better performance by using both spectral and flow ideas.

**Techniques.** The basic idea behind both `MQI` and `Improve` is to maximize the parameter $\alpha$ such that $\alpha \cdot \deg(u)$ units of flow can be concurrently routed in $G$ from each vertex $u \in A$ to $V - A$, obeying the unit-capacity constraint on all undirected edges in the original graph.

---

[2]Note that any cut $S \subset V$ either $\frac{\text{vol}(A)}{\text{vol}(V-A)}$-overlaps with $A$, or has its complement $\bar{S}$ $\frac{\text{vol}(A)}{\text{vol}(V-A)}$-overlap with $A$.

[3]We remark here that we have made no attempt in this version of the paper to improve the constants hidden in $O(1/\delta)$ and $O(1/\sigma)$. They can be made arbitrarily close to 1.

However, `MQI` and `Improve` differ in the way demand sinks are distributed among $V - A$. `MQI` essentially collapses $V - A$ to a single point, while `Improve` requires each $v \in V - A$ to be the sink for a fixed amount of demand $\alpha \cdot \deg(v) \cdot \frac{\text{vol}(A)}{\text{vol}(V-A)}$. This choice of demand ensures that the total flow into the graph equals the total demand at the sink.

This restriction on the sink demands is modeled by introducing an augmented graph $G(\alpha)$ containing a super-source $s$ and super-sink $t$ and connecting $s$ to each vertex in $A$ and $t$ to each vertex in $V - A$ with edges of appropriate capacity.

The success or failure to the resulting parameterized $s - t$ maximum flow problem provides evidence of whether there exists a cut of conductance smaller than $\alpha$, or all cuts that $\delta$-overlap with $A$ have conductance larger than $\Omega(\alpha/\delta)$ (see Lemma 3.2 later). The optimal $\alpha$ is obtained by performing a binary search over $\alpha$.

We exploit the same idea, but modify the augmented graph, and in particular the capacities between vertices in $V - A$ and the super-sink $t$, so that it admits a local flow solution that still contains almost all the significant information about low-conductance cuts near $A$. A formal definition of our augmented graph and flow problem and how they compare to those used in `MQI` and `Improve` can be found in Section 3 We also describe an optimization perspective of our algorithm and its relation to `MQI` and `Improve`.

Given the definition of the augmented graph, we develop two algorithms that solve the corresponding flow problem in a local manner. Our first algorithm, `LocalFlow`, uses local and approximate maximum flows, i.e. the flow problem is not solved exactly. The high level idea of `LocalFlow` is to use a modified version of Dinic's algorithm [Din70] that is ensured to run locally. More specifically:

- We design a blocking flow algorithm that runs locally in time $\widetilde{O}(\text{vol}(A)/\sigma)$ rather than $\widetilde{O}(m)$ for our new augmented graph which is now parameterized by $\sigma$. This requires an idea similar to the approximate PageRank random walk [ACL06]: we explore neighbors of a vertex $v$ only when $v$ is "fully visited", i.e. its edge to the super-sink is fully saturated. The latter technique ensures that the volume of the set of explored nodes is never larger than $O(\text{vol}(A)/\sigma)$.

- We use Dinic's algorithm that recursively calls the blocking flow algorithm above, but impose an upper bound $I = O(\frac{\log(\text{vol}(A)/\sigma)}{\alpha})$ on the number of iterations. (Recall that $\alpha$ is the amount of flow that we want to concurrently route from $A$ to $V - A$.) If the exact maxflow is computed in the $I$ iterations, we are done. Otherwise, although we only get an approximate flow, we can recover a cut of conductance at most $O(\alpha)$ by considering all the sweep cuts given by arranging the vertices by their distances to the source in the residual graph.

Our second algorithm, `LocalFlow`<sup>exact</sup> attempts to match Andersen and Lang's guarantee using local but exact maximum flows. The running time of the algorithm will depend on $\text{vol}(A)^{1.5}$ rather than $\text{vol}(A)$, but will not depend on the conductance of the output set $\phi(S)$. As a result, this algorithm is stronger than `LocalFlow` for small conductance values, i.e., when $\phi(S) \ll \frac{1}{\text{vol}(A)^{0.5}}$.

The `LocalFlow`<sup>exact</sup> algorithm is essentially a localized version of Goldberg-Rao's algorithm [GR98], but requires a lot of additional efforts. Recall that Goldberg and Rao define length functions in the residual graph to be binary $\{0, 1\}$ (rather than all 1 like Dinic's algorithms) so that the resulting *admissible graph* becomes cyclic. This adds an extra level of difficulty because, in our augmented graph, undirected length-0 edges are abundant and may easily destroy the locality of the algorithm. We resolve this issue by modifying the length function to be binary only between "well visited" vertices, and equal to 1 elsewhere. This enables us to design a local algorithm to provide the exact $s$-$t$ maximum flow on our augmented graph.

## 1.2 $O(1)$-Approximation to Local Partitioning on Well-Connected Clusters

The main motivation to consider the local cut-improvement in this paper is for its application to local graph partitioning. Existing local graph partitioning algorithms [ST04, ST13, ACL06, AP09, OT12] provide guarantees of the following kind: if there exists a set $B \subset V$ with conductance $\phi(B)$, they guarantee the existence of some large set $B^g \subseteq B$ of "good seeds" within the target cut $B$, such that for any "good seed" vertex $v \in B^g$, the algorithm on input $v$, outputs a set $A$ with conductance $\phi(A) = \widetilde{O}(\sqrt{\phi(B)})$. Those works apply spectral ideas and hence appear to be limited to an output conductance of $O(\sqrt{\phi(B)})$, the same approximation achieved by Cheeger's Inequality.

**The Internal Connectivity of a Cluster.** Recently, Zhu, Lattanzi and Mirrokni [ZLM13] made an important step towards enhancing the performance guarantee of such spectral-based method to match the so-called improved Cheeger's Inequality [KLL+13]. This requires the introduction of an additional parameter, the *internal connectivity* $\mathsf{Conn}(B)$ of the target cluster $B$. This is defined as

$$\mathsf{Conn}(B) \overset{\text{def}}{=} \frac{1}{\tau_{\mathrm{mix}}(B)} \in [0, 1] \ ,$$

where $\tau_{\mathrm{mix}}(B)$ is the mixing time for a random walk on the subgraph induced by $B$. We will formalize the definition of $\tau_{\mathrm{mix}}(B)$ as well as provide alternative definitions to $\mathsf{Conn}(B)$ in Section 2.

All local partitioning algorithms prior to [ZLM13] only assume that $\phi(B)$ is small, i.e., $B$ is poorly connected to $V - B$. Such set $B$, no matter how small $\phi(B)$ is, may be poorly connected or even disconnected inside. This should not happen if $B$ is a "good" ground-truth cluster, and thus motivates them to study $\mathsf{Conn}(B)$.

**Finding Sparsest Cut on Well-Connected Clusters.** By exploiting the assumption on the internal connectivity, Zhu, Lattanzi and Mirrokni show that one can get a true approximation factor $O(\sqrt{\mathsf{Conn}(B)})$ for conductance:

**Theorem 2** ([ZLM13]). *If there exists a non-empty set $B \subset V$ such that $\frac{\mathsf{Conn}(B)}{\phi(B)} \geq \Omega(1)$, then there exists some $B^g \subseteq B$ with $\mathrm{vol}(B^g) \geq \frac{1}{2}\mathrm{vol}(B)$ such that, when choosing a starting vertex $v \in B^g$, the* `PageRank-Nibble` *algorithm outputs a set $A$ with*

1. *$\mathrm{vol}(A - B) \leq O\big(\frac{\phi(B)}{\mathsf{Conn}(B)}\big) \cdot \mathrm{vol}(B)$,*
2. *$\mathrm{vol}(B - A) \leq O\big(\frac{\phi(B)}{\mathsf{Conn}(B)}\big) \cdot \mathrm{vol}(B)$,*
3. *$\phi(A) \leq O\big(\frac{\phi(B)}{\sqrt{\mathsf{Conn}(B)}}\big) = O\big(\sqrt{\frac{\phi(B)}{\mathsf{Conn}(B)/\phi(B)}}\big)$, and*

*with running time $O\big(\frac{\mathrm{vol}(B)}{\mathsf{Conn}(B)}\big) \leq O\big(\frac{\mathrm{vol}(B)}{\phi(B)}\big)$.*

**Remark 1.1.** This result essentially matches the global result on the improved Cheeger's Inequality [KLL+13]. Let $\phi_{\mathsf{opt}}$ be the optimal conductance of $G$, and $v$ the second eigenvector of the normalized Laplacian matrix of $G$. Using Cheeger's Inequality, one can show that the best sweep cut on $v$ provides a conductance of $O(\sqrt{\phi_{\mathsf{opt}}})$; the improved Cheeger's Inequality says that the conductance guarantee can be improved to $O(\frac{\phi_{\mathsf{opt}}}{\sqrt{\lambda_3}})$ where $\lambda_3$ is the third smallest eigenvalue. In other words, the performance (for the same algorithm) is improved when for instance both sides of the desired cut are well-connected (e.g., expanders). Theorem 2 shows that this same behavior occurs for random-walk based local algorithms.

Zhu, Lattanzi and Mirrokni define $\mathsf{Gap}(B) \overset{\text{def}}{=} \frac{\mathsf{Conn}(B)}{\phi(B)}$, and call a cluster $B$ *well-connected* if $\mathsf{Gap}(B) \geq \Omega(1)$, meaning that $B$ is better connected inside than it is connected to $\bar{B}$. This assumption is particularly relevant when the edges of the graph represent pairwise similarity scores extracted from a machine learning algorithm: we would expect similar nodes to be well connected within themselves while dissimilar nodes to be loosely connected. Theorem 2 says that one can outperform the Cheeger-like output conductance of $O(\sqrt{\phi(B)})$ whenever $\mathsf{Gap}(B) \geq \Omega(1)$, and by a factor of $\sqrt{\mathsf{Gap}(B)}$.

Zhu, Lattanzi and Mirrokni also prove that $O(\phi(B)/\sqrt{\mathsf{Conn}(B)})$ is tight among (a reasonable class of) random-walk based local algorithms. This leads to an interesting open question of designing flow-based local algorithms to further improve the approximation guarantee under the well-connected assumption.

**Our Results.** We resolve the above open question by using the flow-based `LocalImprove` approach and obtain a constant approximation to the conductance.

**Theorem 3.** *Under the same assumption as Theorem 2,* `PageRank-Nibble+LocalImprove` *outputs a set $S$ with $\phi(S) \leq O(\phi(B))$ and $\mathrm{vol}(S) \leq O(\mathrm{vol}(B))$, running in time of $\widetilde{O}(\frac{\mathrm{vol}(B)}{\phi(S)})$.*

This theorem follows immediately from Theorem 2 and Theorem 1a. In fact, Theorem 2 suggests that there exists some constant $c$ (that does not depend on the size of the input instance) such that when $\mathsf{Gap}(B) \geq c$, `PageRank-Nibble` can output some set $A$ with $\mathrm{vol}(A - B) \leq \frac{1}{3}\mathrm{vol}(B)$ and $\mathrm{vol}(B - A) \leq \frac{1}{3}\mathrm{vol}(B)$.

As a result, we can let $\sigma = \frac{2}{3}$ and obtain $\frac{\mathrm{vol}(B \cap A)}{\mathrm{vol}(B)} \geq \sigma = \frac{2}{3}$, so that $B$ is $\frac{2}{3}$-overlapping with $A$. This satisfies the assumption of Theorem 1a for letting $S^*$ be our $B$. Therefore, `LocalImprove` can help to obtain another set $S$ with $\phi(S) \leq O(\phi(B))$ and $\mathrm{vol}(S) \leq O(\mathrm{vol}(B)) \leq O(\mathrm{vol}(A))$. The total running time follows from that of Theorem 2 and Theorem 1a.

We remark here that under the same well-connectedness assumption $\mathsf{Gap}(A) \geq \Omega(1)$, Makarychev, Makarychev and Vijayaraghavan [MMV12] use a global `SDP`-based approach to obtain a constant approximation to the balanced separator or small-set expansion problems. Our Theorem 3 can therefore be viewed as a local algorithm that precisely matches their approximation bound without recurring to an `SDP`-based approach.

**Certifying Expansion using Flows.** In their seminal work, Arora, Rao and Vazirani [ARV09] show that one can certify the expansion of a graph $G$ by a flow-routing of a scaled expander in $G$. This idea was exploited by a number of subsequent works to obtain $O(\mathsf{polylog}(n))$-approximation for graph partitioning problems using only a small number of $s-t$ maximum flows through a primal-dual framework based on a cut-matching game [KRV06, OSVV08]. These works implicitly apply the algorithm `Improve` of Andersen and Lang [AL08] to refine cuts found via spectral methods. To construct the required expander-flow certificate, they notice that `Improve` can be seen as providing a certificate of expansion for cuts near $A$ in the form of the demands routed from $A$ to $V - A$. In the worst case, this certificate amounts to the overlapping guarantee of Theorem 1a and Theorem 1b, but it is better for specific cuts that separate many of the routed demands. In Appendix C, we show that our local algorithms provide the same kind of certificate. This opens the question of whether there exist local versions of the algorithms based on the cut-matching game. Such algorithms would be of great interest as no local graph-partitioning algorithm currently achieves a polylogarithmic approximation guarantee to conductance unconditionally.

**Roadmap.** We provide necessary notations in Section 2, and formally define our revised augmented graph in Section 3. We provide the proof of Theorem 1a in Section 4; we provide the proof of Theorem 1b in Appendix B. We also restate our theorems in their stronger forms using flow certificates in Appendix C, and summarize useful notations in Table 1 on page 20.

## 2 Notation

Consider an undirected graph $G(V, E)$ with $n = |V|$ vertices and $m = |E|$ edges. For any vertex $u \in V$ the degree of $u$ is denoted by $\deg(u)$, and for any subset of the vertices $S \subseteq V$, *volume* of $S$ is denoted by $\mathrm{vol}(S) \overset{\text{def}}{=} \sum_{u \in S} \deg(u)$. Given two subsets $A, B \subset V$, let $E(A, B)$ be the set of edges between $A$ and $B$, let $N(A)$ be the vertices that are adjacent to $A$, and let $\partial A = E(A, N(A))$ be the set of edges on the boundary of $A$.

For a vertex set $S \subseteq V$, we denote by $G[S]$ the induced subgraph of $G$ on $S$ with outgoing edges removed, by $\deg_S(u)$ the degree of $u \in S$ in $G[S]$, and by $\mathrm{vol}_S(T)$ the volume of $T \subseteq S$ in $G[S]$.

We define the *(cut) conductance* and the *set conductance* of a non-empty set $S \subseteq V$ as:

$$\phi(S) \overset{\text{def}}{=} \frac{|E(S, \bar{S})|}{\min\{\mathrm{vol}(S), \mathrm{vol}(\bar{S})\}} \quad \text{and}$$

$$\phi_{\mathsf{s}}(S) \overset{\text{def}}{=} \min_{\varnothing \subset T \subset S} \frac{|E(T, S - T)|}{\min\{\mathrm{vol}_S(T), \mathrm{vol}_S(S - T)\}} \ .$$

Here $\phi_{\mathsf{s}}(S)$ is classically known as the conductance of $S$ on the induced subgraph $G[S]$.

**Well-Connectedness Assumption.** We say a set $B$ is *well-connected* if it satisfies the following gap assumption:

$$\mathsf{Gap}(B) \overset{\text{def}}{=} \frac{\mathsf{Conn}(B)}{\phi(B)} \overset{\text{def}}{=} \frac{1/\tau_{\mathrm{mix}}(B)}{\phi(B)} \geq \Omega(1) \ ,$$

where $\tau_{\mathrm{mix}}(B)$ is the *mixing time for the relative pointwise distance* in $G[B]$.[4] This assumption can be understood as the cluster $B$ is more well-connected inside than it is connected to $V - B$.

For all results in this paper as well as [ZLM13], one can replace the definition of $\mathsf{Conn}(B)$ with $\mathsf{Conn}(B) \overset{\text{def}}{=} \frac{\phi_{\mathsf{s}}(B)^2}{\log \mathrm{vol}(B)}$, or $\mathsf{Conn}(B) \overset{\text{def}}{=} \frac{\lambda(B)}{\log \mathrm{vol}(B)}$ where $\lambda(B)$ is the *spectral gap*, i.e., 1 minus the second largest eigenvalue of the random walk matrix on $G[B]$. We refer interested readers to a discussion in [ZLM13].

**Residual Graphs.** Given a capacitated graph $G'$ along with an *s-t* flow $f$ respecting the capacities, we denote by $G'_f$ the residual graph. For details about our notions of residual graph, as well as self-contained descriptions of Dinic's and Goldberg-Rao's maximum flow algorithms, see Appendix A.

## 3 Our Modification of Andersen and Lang's Augmented Graph

In order to improve the conductance of a given set $A \subset V$, Andersen and Lang [AL08] proposed to study the *s-t* maximum flow problem on an augmented graph related to $G$ parameterized by $\alpha$. We revise their definition by parameterizing the augmented graph with two parameters, $\alpha$ and $\varepsilon$:

**Definition 3.1** (Augmented Graph). *Given an undirected graph $G = (V, E)$, a vertex subset $A \subset V$ satisfying $\mathrm{vol}(A) \leq \mathrm{vol}(V - A)$, a parameter $\alpha \in (0, 1]$ and a parameter $\varepsilon \in \left[\frac{\mathrm{vol}(A)}{\mathrm{vol}(V - A)}, \infty\right)$, we define the augmented graph $G_A(\alpha, \varepsilon)$ as the capacitated directed graph satisfying (see Figure 1):*

- *The vertex set of $G_A(\alpha, \varepsilon)$ contains all vertices in $V$ plus two special ones, the super-source vertex $s$ and the super-sink vertex $t$.*

---

[4]See for instance [MR95, Definition 6.14]), that is, the minimum time required for a lazy random walk to mix *relatively* on all vertices regardless of the starting distribution. Formally, let $W_B$ be the lazy random walk matrix on $G[B]$, and $\pi$ be the stationary distribution on $G[B]$ that is $\pi(u) = \deg_B(u)/\mathrm{vol}_B(B)$, then $\tau_{\mathrm{mix}} = \min\left\{t \in \mathbb{Z}_{\geq 0} : \max_{u,v}\left|\frac{(\chi_v W_B^t)(u) - \pi(u)}{\pi(u)}\right| \leq \frac{1}{2}\right\}$.

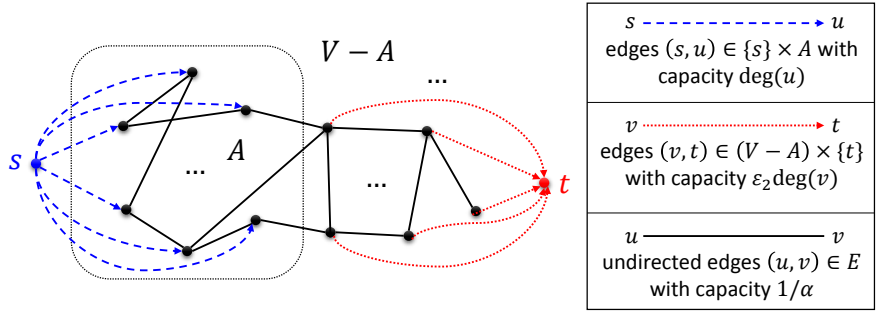

Figure 1: The augmented graph $G_A(\alpha, \varepsilon)$.

- *The edge set of $G_A(\alpha, \varepsilon)$ contains all original undirected edges $(u, v) \in E$ each with capacity $1/\alpha$; in addition, for each $u \in A$ we add a directed edge $s$ to $u$ with capacity $\deg(u)$, and for each $v \in V - A$, we add a directed edge $v$ to $t$ with capacity $\varepsilon \deg(v)$.*

We will be interested in $s$-$t$ cuts on this graph $G_A(\alpha, \varepsilon)$. Recall that an *$s$-$t$ cut* is a set of edges that separates $s$ from $t$, and often denoted by $C$. In our paper, we very often denote an $s$-$t$ cut on this augmented graph $G'$ by a set of vertices $S \subset V$ from the original graph, where one should understand it as a cut $C = \big\{(u, v) \in E : u \in \{s\} \cup S, \ v \in \{t\} \cup (V - S)\big\}$.

For the application in [AL08], they have chosen $\varepsilon$ to be a precise constant $\frac{\mathrm{vol}(A)}{\mathrm{vol}(V-A)}$. For the purpose of this paper, it is crucial to choose larger $\varepsilon$, as we can ensure that any $s$-$t$ maximum flow on $G_A(\alpha, \varepsilon)$ has a local solution, i.e., a solution whose support has volume at most $O((1+1/\varepsilon)\mathrm{vol}(A))$. In fact, given parameter $\sigma$, we will always choose $\varepsilon = \varepsilon_\sigma \overset{\text{def}}{=} \frac{1}{3(1/\sigma - 1)}$ as a function of $\sigma$.

We now state a generalization of the core lemma from Andersen and Lang [AL08] as two statements. The first one says if the max-flow-min-cut value of the augmented graph $G' = G_A(\alpha, \varepsilon_\sigma)$ is smaller than $\mathrm{vol}(A)$, then any such cut $S$ also has a small conductance $\phi(S) < \alpha$. The second one says if the max-flow-min-cut value of the augmented graph $G' = G_A(\alpha, \varepsilon_\sigma)$ is $\mathrm{vol}(A)$, then all sets $S^*$ that are $\sigma$-overlapping to $A$ must have large conductance. The lemma makes it obvious why finding small $s$-$t$ cuts in $G'$ is an interesting question to explore in terms of finding clusters with small conductance.

**Lemma 3.2.** *In the augmented graph $G' \overset{\text{def}}{=} G_A(\alpha, \varepsilon_\sigma)$, where $\varepsilon_\sigma = \frac{1}{3(1/\sigma - 1)} \in \big[\frac{\mathrm{vol}(A)}{\mathrm{vol}(V-A)}, \infty\big)$*

1. *(cut certificate) if there exists an $s$-$t$ cut $S$ with value $\mathtt{cut\text{-}value}_{G'}(S) < \mathrm{vol}(A)$, then $\phi(S) < \alpha$.*

2. *(flow certificate) if there exists an $s$-$t$ flow $f$ with value $\mathtt{flow\text{-}value}_{G'}(f) = \mathrm{vol}(A)$,[5] then for all non-empty subset $S \subseteq V$ we have*

$$\frac{|E(S, V - S)|}{\mathrm{vol}(S)} \geq \alpha \left( \frac{\mathrm{vol}(A \cap S)}{\mathrm{vol}(S)} - \varepsilon_\sigma \frac{\mathrm{vol}(S - A)}{\mathrm{vol}(S)} \right)$$

$$= \alpha \left( (1 + \varepsilon_\sigma) \frac{\mathrm{vol}(A \cap S)}{\mathrm{vol}(S)} - \varepsilon_\sigma \right) \ .$$

*The above guarantee implies that for all $S$ that $\delta$-overlap with $A$ for $\delta \geq \sigma$:*

$$\frac{|E(S, V - S)|}{\mathrm{vol}(S)} \geq \frac{2\alpha}{3} \frac{\mathrm{vol}(A \cap S)}{\mathrm{vol}(S)} \geq \frac{2\alpha}{3} \delta \ .$$

---

[5]We remark here that the flow value cannot exceed $\mathrm{vol}(A)$ because the total amount of capacity going out of $s$ in any augmented graph is only $\mathrm{vol}(A)$.

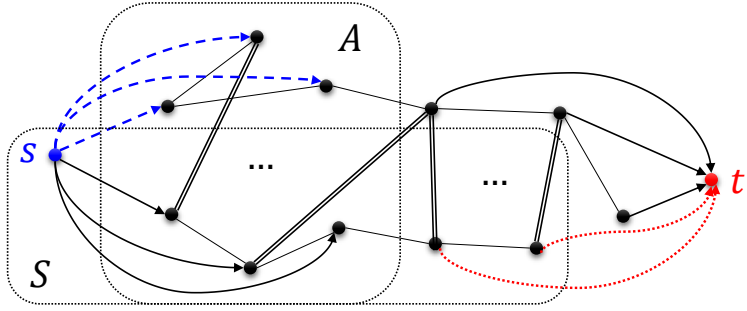

Figure 2: `cut-value`$_{G'}(S)$ consists of three parts: a value of $\mathrm{vol}(A-S)$ for the (blue) dashed edges, $\mathrm{vol}(S-A)$ for the (red) dotted edges, and $\frac{|E(S,V-S)|}{\alpha}$ for the (black) doubled edges. See (3.1).

(We defer the proof to Section 3.1.)

**An Optimization View.** Here, we give a brief motivation of our choice of modified augmented graph through an optimization interpretation. Like Andersen and Lang, the optimization problem we solve is try to the minimization of $\alpha$ such that the the $s$-$t$ maximum flow problem on $G_A(\alpha, \varepsilon_\sigma)$ has a value of $\mathrm{vol}(A)$. It is possible to show that, for the choice of $\varepsilon_\sigma = \frac{1}{3(1/\sigma-1)}$, the above minimization problem over $\alpha$ is equivalent to the following optimization program, where $C \in \left(\frac{\sigma}{3}, 1\right]$ implicitly depends on $\varepsilon_\sigma$:

$$\text{minimize}_{S \subseteq V} \quad \frac{|E(S,V-S)|}{\mathrm{vol}(S \cap A) - \mathrm{vol}(S-A)\frac{\mathrm{vol}(A)}{\mathrm{vol}(V-A)}}$$

$$\text{subject to} \quad \frac{\mathrm{vol}(S \cap A)}{\mathrm{vol}(S)} \geq C \ .$$

Notice that, the objective of this optimization problem is exactly the "quotient score relative to $A$" studied in Andersen and Lang [AL08]. Hence, our algorithm can be seen as optimizing a locally constrained version of the same objective. The presence of this local constraint in the formulation of our problem is what ultimately allows us to construct a local algorithm.

Finally, it should be remarked that this approach to localization is entirely analogous to the one taking place for local random walks. In particular, it has been show that the Personalized PageRank random walk, which is the most commonly used local random walk, can be interpreted as a locally constrained version of the minimum eigenvector problem[MOV12].

## 3.1 Proof of Lemma 3.2

For any $s$-$t$ cut $S$ in $G'$, we rewrite its cut value as follows. According to Figure 2:[6]

$$\texttt{cut-value}_{G'}(S) = \frac{|E(S,V-S)|}{\alpha} + \mathrm{vol}(A-S) + \varepsilon_\sigma \mathrm{vol}(S-A)$$

$$= \mathrm{vol}(A) - \left(\mathrm{vol}(A \cap S) - \varepsilon_\sigma \mathrm{vol}(S-A) - \frac{|E(S,V-S)|}{\alpha}\right) \ . \quad (3.1)$$

We prove the cut certificate part first. Since $\texttt{cut-value}_{G'}(S) < \mathrm{vol}(A)$ for the given cut, this $S$ cannot equal to $\varnothing$ or $V$ as otherwise its cut-value would precisely equal to $\mathrm{vol}(A)$. We deduce

---

[6]We slightly abuse the notation and use $\texttt{cut-value}(S)$ where $S$ is a set to denote $\texttt{cut-value}(C)$ where $C$ is a cut.

from the above formula

$$\mathrm{vol}(A) - \left(\mathrm{vol}(A \cap S) - \varepsilon_\sigma \mathrm{vol}(S - A) - \frac{|E(S, V - S)|}{\alpha}\right) < \mathrm{vol}(A)$$

$$\Longleftrightarrow |E(S, V - S)| < \alpha \left(\mathrm{vol}(A \cap S) - \varepsilon_\sigma \mathrm{vol}(S - A)\right) \ .$$

This implies two things. First,

$$\Longrightarrow \frac{|E(S, V - S)|}{\mathrm{vol}(S)} < \alpha \left(\frac{\mathrm{vol}(A \cap S)}{\mathrm{vol}(S)} - \varepsilon_\sigma \frac{\mathrm{vol}(S - A)}{\mathrm{vol}(S)}\right) \le \alpha \ .$$

Second,

$$\Longrightarrow \frac{|E(S, V - S)|}{\mathrm{vol}(V - S)} < \alpha \left(\frac{\mathrm{vol}(A \cap S)}{\mathrm{vol}(V - S)} - \varepsilon_\sigma \frac{\mathrm{vol}(S - A)}{\mathrm{vol}(V - S)}\right)$$

$$= \alpha \left(\frac{\mathrm{vol}(A) - \mathrm{vol}(A \cap (V - S))}{\mathrm{vol}(V - S)} - \varepsilon_\sigma \frac{\mathrm{vol}(V - A) - \mathrm{vol}((V - S) - A)}{\mathrm{vol}(V - S)}\right)$$

$$\le \alpha \left(\frac{-\mathrm{vol}(A \cap (V - S))}{\mathrm{vol}(V - S)} - \varepsilon_\sigma \frac{-\mathrm{vol}((V - S) - A)}{\mathrm{vol}(V - S)}\right)$$

$$\text{(using } \varepsilon_\sigma \ge \tfrac{\mathrm{vol}(A)}{\mathrm{vol}(V - A)} \ge \tfrac{\mathrm{vol}(A)}{\mathrm{vol}(V - A)})$$

$$\le \alpha \left(\varepsilon_\sigma \frac{\mathrm{vol}((V - S) - A)}{\mathrm{vol}(V - S)}\right) \le \alpha \ .$$

In sum, we have $\phi(S) < \alpha$ by the definition of cut conductance.

Now we show the flow certificate part. Using (3.1) again and the fact that for any non-empty set $S \subseteq V$ we have $\texttt{cut-value}_{G'}(S) \ge \texttt{flow-value}_{G'}(f) = \mathrm{vol}(A)$, we obtain that

$$\mathrm{vol}(A \cap S) - \varepsilon_\sigma \mathrm{vol}(S - A) - \frac{|E(S, V - S)|}{\alpha} \le 0 \ ,$$

and this immediately implies the desired result after multiplying $\frac{\alpha}{\mathrm{vol}(S)}$ on both sides. $\qquad \square$

# 4 Local Improvement Using Approximate Maximum Flow

In this section we prove Theorem 1a by considering a local but approximate maximum flow algorithm for the augmented graph $G' \stackrel{\text{def}}{=} G_A(\alpha, \varepsilon_\sigma)$ introduced in Section 3.

In particular, our proof can be divided into three steps. In the first step, we propose a local algorithm for computing blocking flows in $G'$ whose running time is nearly-linear $\widetilde{O}(\mathrm{vol}(A))$ in terms of the size of $A$. In the second step, we implement Dinic's algorithm but force it to stop after $\Theta(\frac{\log \mathrm{vol}(A)}{\alpha})$ iterations. If this early termination provides an exact $s$-$t$ maximum flow in $G'$ then we can use Lemma 3.2 to produce a flow/cut certificate; otherwise, the flow computed may only be approximate,[7] but nevertheless if one studies all the "layered cuts" in the residual graph, one of them enjoys a good conductance. In the third step, we put all things together and perform a binary search on $\alpha$.

---

[7]In fact, this flow is an additive approximation to the maximum flow, and one can use this to deduce an alternative local algorithm for our purpose, leading to a slightly weaker conductance guarantee but faster by a logarithmic factor.

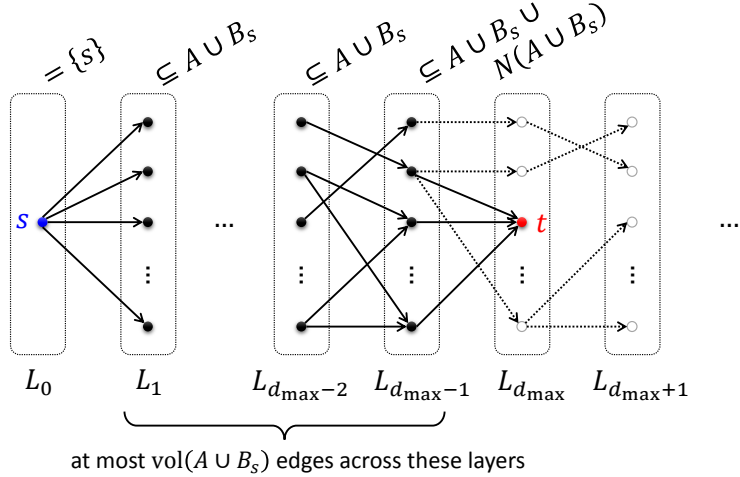

Figure 3: The layers of the admissible graph for a blocking flow problem. The set inclusions follow from Lemma 4.1. The solid edges and vertices are part of the local graph in Definition 4.2.

## 4.1 Localizing Blocking Flows

We now look at the problem of computing blocking flow in the residual graph $G'_f$ for some given flow $f$.

Let $d(v)$ be the shortest path distance from $s$ to $v$ in the residual graph of $G'_f$. Thus, we always have $d(s) = 0$, $d(t) \geq 3$ by our construction of the graph $G' = G_A(\alpha, \varepsilon_\sigma)$. We also denote by $L_j = \{v \in G' \mid d(v) = j\}$ the $j$-th *layer* of this shortest path graph, and $d_{\max} \stackrel{\text{def}}{=} d(t)$ the distance between the super-source and the super-sink. Recall from Dinic's algorithm (cf. Appendix A.1) that the *admissible graph* consists of residual edges $(u, v)$ satisfying $d(u) + 1 = d(v)$, which must be the edges across consecutive layers (see Figure 3), and the *blocking flow problem* aims to find an augmenting flow in this layered admissible graph so that $s$ and $t$ become disconnected.

It is easy to see that any blocking flow algorithm only needs to consider vertices in $L_j$ for $0 \leq j < d_{\max}$, because the distance to the sink $t$ is precisely $d_{\max}$ and there is no need to look at those vertices $v$ whose distance $d(v)$ is as large as $d(t) = d_{\max}$. With this intuition in mind, if we can upper bound the size of $L_j$ for all $j < d_{\max}$ by a value that only depends on $\text{vol}(A)$, we will be able to construct a local blocking flow algorithm.

Indeed, we can upper bound the size of sets $L_j$ with the help from an auxiliary subset, the *saturated set* $B_s \subseteq V$, defined to be the set of vertices whose edges to the super-sink are already fully saturated by $f$ before the blocking flow computation. Under this definition, we provide the following lemma to relate the layer sets to the saturated set and its neighbors.

**Lemma 4.1.** *For every $i \geq 0$, if $d_{\max} < \infty$:*
- *$L_j = \{s\}$ for $j = 0$.*
- *$L_j \subseteq A \cup B_s$ for every $1 \leq j \leq d_{\max} - 2$.*
- *$L_j \subseteq A \cup B_s \cup N(A \cup B_s)$ for $j = d_{\max} - 1$.*

*Proof.* We only need to verify the last two items. We recall that $d_{\max} = d(t)$ is always no smaller than 3 by our construction of the graph $G'$.

For every $1 \leq j \leq d_{\max} - 2$ and an arbitrary $v \in L_j$, we need to show that $v \in A \cup B_s$. If $v \in A$ then we are done, so let us focus on the case when $v \in V - A$ so there is an edge from $v$ to $t$ in $G'$ of capacity $\varepsilon_\sigma \deg(v) > 0$ by the definition of $G'$. In fact, because $d(v) = j \leq d(t) - 2$, it must be true that this edge $(v, t)$ is already saturated, and therefore $v \in B_s$. This proves the second item.

For $j = d_{\max} - 1$ and an arbitrary $v \in L_j$, we need to show that $v \in A \cup B_s \cup N(A \cup B_s)$. Let $(u, v) \in E$ be an arbitrary admissible edge in the residual graph satisfying $d(u) + 1 = d(v)$, then $u \in L_{j-1}$ and $j - 1 \geq 1$. By the previous case, we have that $u \in A \cup B_s$. This concludes that $v$ is a neighbor to some vertex in $A \cup B_s$, and therefore $v \in A \cup B_s \cup N(A \cup B_s)$, proving the third item. $\qquad \square$

We are now ready to state how to make a blocking flow computation *local* on our specific augmented graph $G' = G_A(\alpha, \varepsilon_\sigma)$. In fact, using Lemma 4.1 we can define a local graph $G''$ that is a subgraph of $G'$, and show that the blocking flow computation can be restricted to $G''$.

**Definition 4.2** (local graph). *Given $G' = G_A(\alpha, \varepsilon_\sigma)$ and $B_s$ the saturated set, we define the* local graph $G'' = G'\langle B_s \rangle$ *as a subgraph of $G'$ with (see solid vertices and edges in Figure 3)*

- *vertices $\{s, t\} \cup A \cup B_s \cup N(A \cup B_s)$,*
- *edges $(s, v)$ for all $v \in A$,*
- *edges $(v, t)$ for all $v \in B_s \cup N(A \cup B_s)$, and*
- *edges $(u, v)$ for all $u \in A \cup B_s$ and $v \in V$.*

Notice that the number of edges in $G'' = G'\langle B_s \rangle$ is upper bounded by $O(\mathrm{vol}(A \cup B_s))$. Using the previous fact and the fact that a blocking flow can be computed in time $O(m \log m)$, where $m$ is the number of edges in a graph (cf. Appendix A.1), we can bound the running time of our local blocking flow computation. The proof of the following lemma is routinary, requiring some careful case analyses, and can be found in Section 4.4.

**Lemma 4.3.** *When computing $\mathtt{BlockFlow}_{G', f}(s, t)$, it suffices to compute $\mathtt{BlockFlow}_{G'', f}(s, t)$ on the local graph $G'' = G'\langle B_s \rangle$. Therefore, the running time is $O\big(\mathrm{vol}(A \cup B_s) \log \mathrm{vol}(A \cup B_s)\big)$.*

At last, the next simple lemma gives an upper bound on the size of the saturated set $B_s$, therefore showing that the blocking flow computation of $\mathtt{BlockFlow}_{G'', f}(s, t)$ is local.

**Lemma 4.4.** *We always have $\mathrm{vol}(B_s) \leq \mathrm{vol}(A)/\varepsilon_\sigma$, so $\mathrm{vol}(A \cup B_s) \leq (1 + \frac{1}{\varepsilon_\sigma})\mathrm{vol}(A) < \frac{3}{\sigma}\mathrm{vol}(A)$.*

*Proof.* For any vertex $v \in B_s$, it must satisfy that its edge to the sink $t$ in $G'$ is fully saturated (with a capacity of $\varepsilon_\sigma \deg(v)$) before it is contained in $B_s$. Since the maximum flow value in $G'$ is upper bounded by $\mathrm{vol}(A)$, the total volume of such vertices cannot exceed $\mathrm{vol}(A)/\varepsilon_\sigma$. $\qquad \square$

## 4.2 Localizing Dinic's Algorithm with Early Termination

We are now ready to state our approximate maximum flow algorithm $\mathtt{LocalFlow}$. It is a truncated version of Dinic's algorithm where the maximum number of iterations is $I \stackrel{\text{def}}{=} \lceil \frac{5}{\alpha} \log(3\mathrm{vol}(A)/\sigma) \rceil$ and each iteration of the blocking flow computation can be implemented to run on our local graph $G'' = G'\langle B_s \rangle$ owing to Lemma 4.3. We summarize the pseudocode in Algorithm 1.

We emphasize here that the saturated set $B_s$ can be maintained constructively (see Line 8 in Algorithm 1). Throughout Dinic's algorithm (and in fact any augmenting-path-based maximum flow algorithm), the flow on the edges from vertices $v \in V - A$ to the sink $t$ in $G'$ can never decrease, and therefore, vertices can only enter $B_s$ but never leave it.

If the flow augmentation finishes in $I$ iterations, we get an exact maximum flow $f$ at the end and output its $s$-$t$ mincut. Otherwise, when the maximum number of iterations is reached and the flow augmentation has not finished, we only obtain an approximate flow $f$ whose value

---

**Algorithm 1** $\texttt{LocalFlow}_G(A, \alpha, \varepsilon_\sigma)$

---

**Input:** $G = (V, E)$, $A \subset V$, $\alpha \in (0, 1]$, and $\varepsilon_\sigma \in \left[\frac{\text{vol}(A)}{\text{vol}(V-A)}, \infty\right)$ .

**Output:** an $s$-$t$ flow $f$ and a set $S$ (representing an $s$-$t$ cut $(\{s\} \cup S, \{t\} \cup (V - S))$) in $G_A(\alpha, \varepsilon_\sigma)$.

1: $G' \leftarrow G_A(\alpha, \varepsilon_\sigma)$.                                                     {see Definition 3.1.}

2: $B_s \leftarrow \varnothing$.

3: $f \leftarrow 0$.

4: **for** $i \leftarrow 1$ **to** $I \overset{\text{def}}{=} \lceil \frac{5}{\alpha} \log(3\text{vol}(A)/\sigma) \rceil$ **do**

5:      $G'' \leftarrow G' \langle B_s \rangle$.                                                  {see Definition 4.2.}

6:      $f \leftarrow f + \texttt{BlockFlow}_{G'',f}(s, t)$ and breaks if $\texttt{BlockFlow}$ fails to augment.

7:      $C \leftarrow$ the vertices in $N(A \cup B_s)$ whose edges to the sink get saturated in the new flow

8:      $B_s \leftarrow B_s \cup C$.

9: **end for**

10: **if** $\texttt{BlockFlow}$ ever fails to find an augmenting path **then**

11:      $f$ is now an exact $s$-$t$ maximum flow in $G'$ and let its min-cut be $S$.

12: **else**

13:      $S \leftarrow$ the cut among all layer cuts (w.r.t. the distance labeling in $G'_f$) that minimizes conductance.

14: **end if**

15: **return** $(f, S)$.

---

is strictly smaller than $\text{vol}(A)$. In this case we consider all the cuts between consecutive layers $L_1, L_2, \ldots, L_{d_{\max}-1}$ at the end of the execution of $\texttt{LocalFlow}$. For any $j \in \{1, 2, \ldots, d_{\max} - 2\}$, we denote by $S_j \overset{\text{def}}{=} \bigcup_{k=1}^{j} L_k \subseteq V$ the union of the first $j$ layers. Each such $S_j$ is called a *layer cut* and we output the one among them that minimizes $\phi(S_j)$.

We now provide the key lemma on the performance of this $\texttt{LocalFlow}$ algorithm.

**Lemma 4.5** ($\texttt{LocalFlow}$ performance). *Given arbitrary* $\alpha \in (0, 1]$, $\varepsilon_\sigma \in \left[\frac{\text{vol}(A)}{\text{vol}(V-A)}, \infty\right)$, *and suppose there is some set* $S^*$ *satisfying:*

$$\frac{|E(S^*, V - S^*)|}{\text{vol}(S^*)} < \alpha \left( \frac{\text{vol}(A \cap S^*)}{\text{vol}(S^*)} - \varepsilon_\sigma \frac{\text{vol}(S^* - A)}{\text{vol}(S^*)} \right) \quad . \tag{4.1}$$

*Then,* $\texttt{LocalFlow}_G(A, \alpha, \varepsilon_\sigma)$ *outputs a flow* $f$ *with value strictly smaller than* $\text{vol}(A)$, *and produces a set* $S$ *of size* $\text{vol}(S) \leq \text{vol}(A)(1 + 1/\varepsilon_\sigma)$ *with conductance*

$$\phi(S) = \frac{|E(S, V - S)|}{\min\{\text{vol}(S), \text{vol}(V - S)\}} < 2\alpha \quad .$$

*The running time of* $\texttt{LocalFlow}_G(A, \alpha, \varepsilon_\sigma)$ *is* $O\left( \frac{\text{vol}(A)}{\sigma \alpha} \log^2 \left( \frac{\text{vol}(A)}{\sigma} \right) \right)$.

The proof of the above lemma is somewhat involved and can be found in Section 4.5, but the intuition behind it is quite clear. Given any value of $\alpha$ and $\varepsilon_\sigma$, if the flow augmentation finishes in $I$ iterations (i.e., we get an exact $s$-$t$ flow), this provides either a flow or cut certificate for Lemma 3.2. If (4.1) is satisfied, we cannot have a flow certificate and therefore must have a cut certificate producing a set of conductance less than $\alpha$.

Otherwise, we get an approximate $s$-$t$ flow.[8] In this case, we cannot apply Lemma 3.2, but we can show that one of the layer cuts provides conductance less than $2\alpha$. Indeed, suppose that all layer cuts have conductance larger than or equal to $2\alpha$, then the "(directed) conductance" of all of those layer cuts in the residual graph $G'_f$ should also be larger than or equal to $\alpha$ due to the design of our augmented graph. Next, using a classical ball-growing argument, we conclude that $s$ and $t$ in $G'_f$ must be at distance at most $\log(A \cup B_s)/\alpha \approx I$ away. This contradicts an important property of Dinic's algorithm: after $I$ iterations of blocking flow computation, $s$ and $t$ must be of distance at least $I$ units apart. This concludes the proof.

---

[8]Note that this flow is additively approximate to the maximum $s$-$t$ flow with a good error bound, although we are not going to use this fact in the proof.

### 4.3 The Final Algorithm

Now we are ready to put everything together and construct our final algorithm on the local improvement. Using the result from Lemma 4.5, one can essentially perform a binary search on the value of $\alpha \in (0, 1]$: if `LocalFlow` returns a flow with value $\text{vol}(A)$ it means the choice of $\alpha$ is too small; otherwise it gives a positive solution $S$ with conductance $\phi(S) < 2\alpha$ and we should continue to search for smaller values of $\alpha$. We summarize this binary search algorithm as `LocalImprove` in Algorithm 2, and show a rigorous bound on its running time and conductance guarantee.

---

**Algorithm 2** $\text{LocalImprove}_G(A, \varepsilon_\sigma, \varepsilon)$

**Input:** $G = (V, E)$, $A \subset V$, $\varepsilon_\sigma \in \left[\frac{\text{vol}(A)}{\text{vol}(V-A)}, \infty\right)$ and $\varepsilon \in (0, 1]$.
**Output:** a non-empty set $S \subset V$ with good conductance.

1: $\alpha_{\min} \leftarrow 0, \alpha_{\max} \leftarrow 1$.
2: **while** $\alpha_{\max} - \alpha_{\min} > \varepsilon \alpha_{\min}$ **do**
3:      $\alpha \leftarrow \frac{1}{2}(\alpha_{\min} + \alpha_{\max})$.
4:      **if** $\text{LocalFlow}_G(A, \alpha, \varepsilon_\sigma)$ returns a flow with value $\text{vol}(A)$ **then**
5:          $\alpha_{\min} = \alpha$.
6:      **else**
7:          $\alpha_{\max} = \alpha$.
8:      **end if**
9: **end while**
10: $(f, S) \leftarrow \text{LocalFlow}_G(A, \alpha_{\max}, \varepsilon_\sigma)$.
11: **return** $S$.

---

**Theorem 4.6.** *Given arbitrary $\varepsilon_\sigma \in \left[\frac{\text{vol}(A)}{\text{vol}(V-A)}, \infty\right)$ and $\varepsilon \in (0, 1]$, and suppose there is some set $S^*$ satisfying*

$$\frac{|E(S^*, V - S^*)|}{\text{vol}(S^*)} < \alpha^* \left(\frac{\text{vol}(A \cap S^*)}{\text{vol}(S^*)} - \varepsilon_\sigma \frac{\text{vol}(S^* - A)}{\text{vol}(S^*)}\right) \ , \qquad \text{(same as (4.1))}$$

*then $\text{LocalImprove}_G(A, \varepsilon_\sigma, \varepsilon)$ outputs a set $S$ of size $\text{vol}(S) \leq \text{vol}(A)(1 + 1/\varepsilon_\sigma)$ with conductance*

$$\phi(S) = \frac{|E(S, V - S)|}{\min\{\text{vol}(S), \text{vol}(V - S)\}} < 2(1 + \varepsilon)\alpha^* \ .$$

*In addition, it runs in time $O\left(\frac{\text{vol}(A)}{\sigma \phi(S)} \log^2\left(\frac{\text{vol}(A)}{\sigma}\right) \log\left(\frac{1}{\varepsilon}\right)\right) = \widetilde{O}\left(\frac{\text{vol}(A)}{\sigma \phi(S)}\right)$.*

*Proof.* The proof for this theorem is routinary.

First, whenever a value $\alpha \geq \alpha^*$ is chosen during the binary search, according to Lemma 4.5, `LocalFlow` should return a set with conductance $\phi(S) < 2\alpha$. In the case when $\alpha < \alpha^*$, `LocalFlow` may fail (i.e., outputting a flow of value $\text{vol}(A)$) or output a set with conductance $\phi(S) < 2\alpha < 2\alpha^*$.

In the latter case, if `LocalFlow` ever outputs a set with conductance $\phi(S) < 2\alpha < 2\alpha^*$, we know automatically that the output for `LocalImprove` has a conductance smaller than $2\alpha^*$. Otherwise, if `LocalFlow` always fails whenever $\alpha < \alpha^*$, we have the guarantee that $\alpha_{\min} < \alpha^* \leq \alpha_{\max}$ throughout the algorithm.

At the end of the while loop, not only `LocalFlow` does not fail for the choice of $\alpha = \alpha_{\max}$, we further know that $\alpha_{\max} \leq (1 + \varepsilon)\alpha_{\min} < (1 + \varepsilon)\alpha^*$. As a result, we have the desired conductance bound $\phi(S) < 2(1 + \varepsilon)\alpha^*$ on the final output $S$.

We now bound the running time for `LocalImprove` carefully.

We first notice that finite arithmetic operations are not an issue for our algorithm, because the blocking flow step runs in strongly polynomial time.[9]

Let $k$ be the first iteration of `LocalImprove` in which `LocalFlow` fails. At that moment $\alpha_{\min} = \frac{1}{2^{k+1}}$ and $\alpha_{\max} = \frac{1}{2^k}$. Letting $\widetilde{\alpha}$ be the conductance of the final output set in `LocalImprove`, we immediately know that $\widetilde{\alpha} \in [\frac{1}{2^{k+1}}, \frac{1}{2^k}]$. Since the value of $\alpha$ decreases by a factor of 2 in the first $k$ iterations, and the running time of `LocalFlow` depends inversely proportional to $\alpha$, we have that the running time for the first $k$ iteration is dominated by its last run, which is $O\left(\frac{\mathrm{vol}(A)}{\sigma\widetilde{\alpha}} \log^2\left(\frac{\mathrm{vol}(A)}{\sigma}\right)\right)$.

Next, because $\alpha_{\max} - \alpha_{\min} = \frac{1}{2^{k+1}} \approx \widetilde{\alpha}$, and our algorithm `LocalImprove` stops when $\alpha_{\max} - \alpha_{\min} \leq \varepsilon\alpha_{\min} \leq \varepsilon\widetilde{\alpha}$, this suggests that there are at most $O\left(\log \frac{1}{\varepsilon}\right)$ more iterations left in the binary search, each taking a running time of $O\left(\frac{\mathrm{vol}(A)}{\sigma\widetilde{\alpha}} \log^2\left(\frac{\mathrm{vol}(A)}{\sigma}\right)\right)$.

In sum, we have shown that the running time for `LocalImprove` is $O\left(\frac{\mathrm{vol}(A)}{\sigma\widetilde{\alpha}} \log^2\left(\frac{\mathrm{vol}(A)}{\sigma}\right) \log \frac{1}{\varepsilon}\right)$, where $\widetilde{\alpha}$ is the conductance for the output set $S$. $\qquad\square$

Now Theorem 1a is a direct corollary of the above (and in fact, more general) theorem.

**Theorem 1a** (restated). *Given set $A \subset V$ of the graph and a constant $\sigma \in (0, 1]$ satisfying $\frac{\mathrm{vol}(V-A)}{\mathrm{vol}(A)} \geq 3(1/\sigma - 1)$, and suppose*

$$\exists \text{ (unknown) target set } S^* \subseteq V \text{ satisfying } \mathrm{vol}(S^*) \leq \mathrm{vol}(V - S^*) \text{ and } \frac{\mathrm{vol}(A \cap S^*)}{\mathrm{vol}(S^*)} = \delta \geq \sigma,$$

*then our `LocalImprove` algorithm can be parameterized by $\sigma$ and outputs a set $S$ satisfying*

$$\mathrm{vol}(S) \leq \frac{3}{\sigma}\mathrm{vol}(A) \quad and \quad \phi(S) \leq \frac{4}{\delta}\phi(S^*)$$

*in time*

$$O\left(\frac{\mathrm{vol}(A)}{\sigma\phi(S)} \log^2\left(\mathrm{vol}(A)/\sigma\right)\right) \ .$$

---

[9]In other words, since the values of $\alpha$ used in `LocalImprove` are rational, and the value of $\varepsilon_\sigma$ can be assumed without loss of generality to be rational as well, the augmented graph $G_A(\alpha, \varepsilon_\sigma)$ can be viewed as having $O(\log \mathrm{vol}(A))$-bit integral capacities after proper scaling. An arithmetic operation on integral values of this size can be done in $O(1)$ time on a unit-cost RAM.

*Proof.* Recall that we have chosen $\varepsilon_\sigma = \frac{1}{3(1/\sigma-1)} \geq \frac{\mathrm{vol}(A)}{\mathrm{vol}(V-A)}$, and now let $\alpha = \phi(S^*) \cdot 5/3\sigma$ we have

$$
\begin{aligned}
\frac{|E(S^*, V - S^*)|}{\mathrm{vol}(S^*)} = \phi(S^*) = \frac{3\alpha\sigma}{5} < \frac{2\alpha\sigma}{3} &\leq \frac{2\alpha}{3} \frac{\mathrm{vol}(A \cap S^*)}{\mathrm{vol}(S^*)} \\
&= \alpha \left( \frac{\mathrm{vol}(A \cap S^*)}{\mathrm{vol}(S^*)} - \frac{1}{3} \frac{\mathrm{vol}(A \cap S^*)}{\mathrm{vol}(S^*)} \right) \\
&\leq \alpha \left( \frac{\mathrm{vol}(A \cap S^*)}{\mathrm{vol}(S^*)} - \varepsilon_\sigma (1/\sigma - 1) \frac{\mathrm{vol}(A \cap S^*)}{\mathrm{vol}(S^*)} \right) \\
&\leq \alpha \left( \frac{\mathrm{vol}(A \cap S^*)}{\mathrm{vol}(S^*)} - \varepsilon_\sigma \frac{\mathrm{vol}(S^*) - \mathrm{vol}(A \cap S^*)}{\mathrm{vol}(S^*)} \right) \\
&= \alpha \left( \frac{\mathrm{vol}(A \cap S^*)}{\mathrm{vol}(S^*)} - \varepsilon_\sigma \frac{\mathrm{vol}(S^* - A)}{\mathrm{vol}(S^*)} \right) \ .
\end{aligned}
$$

This satisfies the requirement of Theorem 4.6, and letting $\varepsilon = 1/5$, $\mathtt{LocalImprove}^{\mathtt{exact}}$ guarantees to output a set $S$ with conductance $\phi(S) \leq 2(1+1/5)\alpha = \frac{4}{\sigma}\phi(S^*)$, and size $\mathrm{vol}(S) \leq \mathrm{vol}(A)(1+1/\varepsilon_\sigma) \leq \frac{3}{\sigma}$. The running time follows by substituting our choices of $\varepsilon_\sigma, \varepsilon$ into Theorem 4.6. $\qquad\square$

### 4.4 Proof of Lemma 4.3

**Lemma 4.3.** *When computing* $\mathtt{BlockFlow}_{G',f}(s,t)$, *it suffices to compute* $\mathtt{BlockFlow}_{G'',f}(s,t)$ *on the local graph* $G'' = G'\langle B_s \rangle$. *Therefore, the running time is* $O(\mathrm{vol}(A \cup B_s) \log \mathrm{vol}(A \cup B_s))$.

*Proof.* We first show that $s$ and $t$ are disconnected in the residual graph $G'_f$ if and only if they are disconnected in $G''_f$. The only interesting direction is to prove that

$$s \text{ and } t \text{ are disconnected in } G'',$$
$$\text{(so that } \mathtt{BlockFlow}_{G'',f}(s,t) \text{ fails to augment)}$$
$$\implies s \text{ and } t \text{ are disconnected in } G'_f,$$
$$\text{(so that } \mathtt{BlockFlow}_{G',f}(s,t) \text{ must also fail to augment)}$$

Indeed, if $s$ and $t$ are disconnected in $G''_f$, let $W$ be a set that separates $s$ and $t$: i.e., $s \in W$, $t \notin W$, and there is no edge in $G''_f$ between $W$ and $\{t\} + V - W$.

It is now easy to see that $W \subseteq A \cup B_s$ because all other vertices in $N(A \cup B_s)$ must have an edge to the sink $t$ with positive residual capacity. However, notice that all edges that are on the boundary of $A \cup B_s$ are included both in $G''$ and $G'$, so $W$ is also a $s$-$t$ separator in the original residual graph $G'_f$, proving that $s$ and $t$ are disconnected in $G'_f$.

Therefore, for the rest of the proof, we can assume $d(t) = d_{\max} < \infty$. Recall that $L_0, L_1, \ldots, L_{d_{\max}}$ are the layers of vertices in the residual graph $G'_f$ that have distance $0, 1, \ldots, d_{\max}$ respectively from the source $s$. Since the sink $t$ has a distance $d(t) = d_{\max}$, there is no need for the blocking flow to ever touch any vertex whose distance from $s$ is *larger than or equal to* $d_{\max}$ except $t$ itself.

This means, the vertices in the support of the blocking flow $\mathtt{BlockFlow}_{G',f}(s,t)$ must be a subset of

$$\{s, t\} \cup L_1 \cup \cdots \cup L_{d_{\max}-1} \subseteq \{s, t\} \cup A \cup B_s \cup N(A \cup B_s) \ ,$$

(see the solid vertices in Figure 3 and recall Lemma 4.1) and are thus in $G''$.

We then turn to the edges in the support of $\mathtt{BlockFlow}_{G',f}(s,t)$. By the definition of blocking flow, those edges must be (see the solid edges in Figure 3 and recall Lemma 4.1):

- from $s$ to $L_1$, which is of the form $\{(s,v) : v \in A\}$;

- from $L_{d_{\max}-1}$ to $t$, which is of the form $\big\{(v,t) : v \in B_s \cup N\big(A \cup B_s\big)\big\}$; or
- from $L_j$ to $L_{j+1}$ for $1 \leq j \leq d_{\max} - 2$, which is of the form $\big\{(u,v) : u \in A \cup B_s, v \in V\big\}$.

In sum, the edges in the support of $\texttt{BlockFlow}_{G',f}(s,t)$ are also in $G''$. This essentially ends the proof because when computing $\texttt{BlockFlow}_{G',f}(s,t)$, it suffices to compute $\texttt{BlockFlow}_{G'',f}(s,t)$ on $G''$ which guarantees to contain the support of the blocking flow.

At last we focus on the running time. The total number of edges in $G''$ is $O\big(\text{vol}(A \cup B_s)\big)$, and therefore the running time for $\texttt{BlockFlow}_{G'',f}(s,t)$ is $O\big(\text{vol}(A \cup B_s) \log \text{vol}(A \cup B_s)\big)$ owing to Appendix A.1. $\qquad\square$

## 4.5 Proof of Lemma 4.5

**Lemma 4.5** ($\texttt{LocalFlow}$ performance)**.** *Given arbitrary $\alpha \in (0,1]$, $\varepsilon_\sigma \in [\frac{\text{vol}(A)}{\text{vol}(V-A)}, \infty)$, and suppose there is some set $S^*$ satisfying:*

$$\frac{|E(S^*, V - S^*)|}{\text{vol}(S^*)} < \alpha \left( \frac{\text{vol}(A \cap S^*)}{\text{vol}(S^*)} - \varepsilon_\sigma \frac{\text{vol}(S^* - A)}{\text{vol}(S^*)} \right) . \tag{4.1}$$

*Then, $\texttt{LocalFlow}_G(A, \alpha, \varepsilon_\sigma)$ outputs a flow $f$ with value strictly smaller than $\text{vol}(A)$, and produces a set $S$ of size $\text{vol}(S) \leq \text{vol}(A)(1 + 1/\varepsilon_\sigma)$ with conductance*

$$\phi(S) = \frac{|E(S, V - S)|}{\min\{\text{vol}(S), \text{vol}(V - S)\}} < 2\alpha .$$

*The running time of $\texttt{LocalFlow}_G(A, \alpha, \varepsilon_\sigma)$ is $O\left(\frac{\text{vol}(A)}{\sigma \alpha} \log^2 \left(\frac{\text{vol}(A)}{\sigma}\right)\right)$.*

*Proof.* Let $G' = G_A(\alpha, \varepsilon_\sigma)$ be the augmented graph as before. There are three cases.

If the maximum flow between $s$ and $t$ is computed in $I$ iterations, and the max-flow-min-cut value is strictly smaller than $\text{vol}(A)$, then the $s$-$t$ min-cut $S$ also has value strictly smaller than $\text{vol}(A)$. According to the cut certificate in Lemma 3.2, we immediately have $\phi(S) < \alpha$ and are done.

If the maximum flow between $s$ and $t$ is computed in $I$ iterations, and the max-flow-min-cut value is precisely $\text{vol}(A)$, in this case the existence of $S^*$ satisfying (4.1) contradicts to the flow certificate in Lemma 3.2 so it cannot happen.

We are only left with the interesting case when $f$ is not yet the maximum $s$-$t$ flow (and thus $I$ iterations are not sufficient in Dinic's algorithm in producing an exact maximum flow). In this case, we want to show that one of the layered cuts must have conductance at most $2\alpha$.

Suppose not, that is, for all layered cuts $S_j \overset{\text{def}}{=} \bigcup_{k=1}^j L_j \subseteq V$ where $j \in \{1, 2, \ldots, d_{\max} - 2\}$, we have $\phi(S_j) \geq 2\alpha$. Let us denote by $C_f(A, B)$ the total residual capacities from vertices in $A$ to vertices in $B$ in the residual graph $G'_f$, where $A \cap B = \varnothing$. Note that the residual graph $G'_f$ is directed, so $C_f(A, B)$ captures only those directed edges from $A$ to $B$ in $G'_f$, and does not necessarily equal to $C_f(B, A)$.

Next we want to show a lower bound on $C_f(L_j, L_{j+1})$. We begin by observing that there cannot be directed edges across more than one layers (see Figure 3), so we have

$$C_f\big(L_j, L_{j+1}\big) = C_f\big(S_j, V - S_j\big) = C_f\big(\{s\} \cup S_j, (V - S_j) \cup \{t\}\big) .$$

Now we denote by $C_0(A, B)$ the total residual capacity in the original graph (with zero flow) from $A$ to $B$, and by straightforward computation we have (cf. (3.1) and Figure 2):

$$C_0\Big(\{s\} \cup S_j, (V - S_j) \cup \{t\}\Big) = \texttt{cut-value}_{G'}(S_j)$$
$$= \frac{|E(S_j, V - S_j)|}{\alpha} + \text{vol}(A - S_j) + \varepsilon_\sigma \text{vol}(S_j - A) \ .$$

Now comes a key observation: the difference between the above two terms, i.e., $C_0\Big(\{s\} \cup S_j, (V - S_j) \cup \{t\}\Big) - C_f\Big(\{s\} \cup S_j, (V - S_j) \cup \{t\}\Big)$, should precisely equal to the total amount of net flow in $f$ that is routed on this cut, and thus is upper bounded by $\text{vol}(A \cap S_j)$ because that is the total amount of capacities from $s$ to $S_j$ in $G'$. Those two combined, implies that

$$C_f\big(L_j, L_{j+1}\big) \geq \frac{|E(S_j, V - S_j)|}{\alpha} + \text{vol}(A - S_j) + \varepsilon_\sigma \text{vol}(S_j - A) - \text{vol}(A \cap S_j) \ .$$

If $\text{vol}(S_j) \leq \text{vol}(V - S_j)$, we have that

$$\frac{\alpha \cdot C_f\big(L_j, L_{j+1}\big)}{\text{vol}(S_j)} \geq \frac{|E(S_j, V - S_j)|}{\text{vol}(S_j)} - \frac{\alpha \cdot \text{vol}(A \cap S_j)}{\text{vol}(S_j)} \geq 2\alpha - \alpha = \alpha \ .$$

If $\text{vol}(S_j) > \text{vol}(V - S_j)$, we have that

$$\frac{\alpha \cdot C_f\big(L_j, L_{j+1}\big)}{\text{vol}(V - S_j)} \geq \frac{|E(S_j, V - S_j)|}{\text{vol}(V - S_j)} + \alpha \cdot \left(\frac{\varepsilon_\sigma \text{vol}(S_j - A)}{\text{vol}(V - S_j)} - \frac{\text{vol}(A \cap S_j)}{\text{vol}(V - S_j)}\right) \geq 2\alpha - \alpha = \alpha \ ,$$

where the last inequality is because

$$\alpha \left(\frac{\text{vol}(A \cap S_j)}{\text{vol}(V - S_j)} - \varepsilon_\sigma \frac{\text{vol}(S_j - A)}{\text{vol}(V - S_j)}\right)$$
$$= \alpha \left(\frac{\text{vol}(A) - \text{vol}(A \cap (V - S_j))}{\text{vol}(V - S_j)} - \varepsilon_\sigma \frac{\text{vol}(V - A) - \text{vol}((V - S_j) - A)}{\text{vol}(V - S_j)}\right)$$
$$\leq \alpha \left(\frac{-\text{vol}(A \cap (V - S_j))}{\text{vol}(V - S_j)} - \varepsilon_\sigma \frac{-\text{vol}((V - S_j) - A)}{\text{vol}(V - S_j)}\right) \qquad \text{(using } \varepsilon_\sigma \geq \frac{\text{vol}(A)}{\text{vol}(V - A)} \geq \frac{\text{vol}(A)}{\text{vol}(V - A)} \text{)}$$
$$\leq \alpha \left(\varepsilon_\sigma \frac{\text{vol}((V - S_j) - A)}{\text{vol}(V - S_j)}\right) \leq \alpha \ .$$

Notice that in both cases we have

$$\frac{\alpha \cdot C_f\big(L_j, L_{j+1}\big)}{\min\{\text{vol}(S_j), \text{vol}(V - S_j)\}} \geq \alpha \ . \tag{4.2}$$

We remark here that (4.2) can be viewed as the fact that the (directed) conductance for all layered cuts are larger than or equal to $\alpha$ (where the $\alpha$ factor on the left hand side is since all edges in $G$ are equipped with a capacity of $1/\alpha$). This implies the diameter of the residual graph $G'_f$ cannot be larger than $I \approx \log \text{vol}(A)/\alpha$, as we will show next.

Recall that $d_{\max} = d(t) \geq 3$ at the beginning of the algorithm and it increases at least by 1 after each blocking flow computation (see Proposition A.1). This implies that $d_{\max} \geq I + 3$ at the end of the execution. We will show that (4.2) contradicts with the fact that $d_{\max}$, the distance between $s$ and $t$ in $G'_f$, is more than $I + 3$.

Let $j^* \in \{1, 2, \ldots, d_{\max} - 2\}$ be the largest index of the layer satisfying $\mathrm{vol}(S_{j^*}) \leq \frac{\mathrm{vol}(V)}{2}$. For each $j \in \{1, 2, \ldots, j^*\}$, we have

$$\mathrm{vol}(L_{j+1}) \geq \frac{\alpha}{2} \cdot C_f(L_j, L_{j+1}) \geq \frac{\alpha}{2}\mathrm{vol}(S_j) = \frac{\alpha}{2}\sum_{k=1}^{j}\mathrm{vol}(L_k) \ .$$

Here the first inequality is because each vertex $v \in L_{j+1}$ with degree $\deg(v)$ can have at most a residual capacity of $\frac{2\deg(v)}{\alpha}$ going from $L_j$ to $v$ in $L_{j+1}$ (where the factor 2 is due to backward arcs in the residual graph $G_f'$). The second inequality uses (4.2). This inequality further implies that $j^* < \lceil I/2 \rceil$ because otherwise we must have

$$\sum_{k=1}^{j^*+1}\mathrm{vol}(L_k) \geq \left(1 + \frac{\alpha}{2}\right)^{I/2}\mathrm{vol}(L_1) \geq \left(1 + \frac{\alpha}{2}\right)^{I/2} \geq \left(1 + \frac{\alpha}{2}\right)^{\frac{2.5}{\alpha}\log(3\mathrm{vol}(A)/\sigma)}$$

$$\overset{\text{Lemma 4.4}}{\geq} \left(1 + \frac{\alpha}{2}\right)^{\frac{2.5}{\alpha}\log(\mathrm{vol}(A \cup B_s))} > e^{\log(\mathrm{vol}(A \cup B_s))} = \mathrm{vol}(A \cup B_s)$$

getting a contradiction to the fact that $L_1, \ldots, L_{j^*} \subseteq A \cup B_s$ from Lemma 4.1.

We now repeat the same analysis but in the backward direction. We first observe that $\{j^* + 1, \ldots, j^* + \lceil I/2 \rceil\} \subseteq \{j^* + 1, \ldots, d_{\max} - 2\}$ since $d_{\max} \geq I + 3 \geq (j^* + \lceil I/2 \rceil) + 2$. Now for any $j \in \{j^* + 1, \ldots, j^* + \lceil I/2 \rceil\}$, we have $\mathrm{vol}(S_j) > \mathrm{vol}(V - S_j)$ by the definition of $j^*$, and therefore

$$\mathrm{vol}(L_j) \geq \frac{\alpha}{2} \cdot C_f(L_j, L_{j+1}) \geq \frac{\alpha}{2}\mathrm{vol}(V - S_j) \geq \frac{\alpha}{2}\sum_{k=j+1}^{d_{\max}-1}\mathrm{vol}(L_k) \ .$$

Here the first inequality is because each vertex $v \in L_j$ with degree $\deg(v)$ can have at most a residual capacity of $\frac{2\deg(v)}{\alpha}$ going from $v$ to $L_{j+1}$ (where the factor 2 is due to backward arcs in the residual graph $G_f'$). The second inequality uses (4.2).

This inequality further implies that

$$\sum_{k=j^*+1}^{j^*+\lceil I/2 \rceil}\mathrm{vol}(L_k) \geq \left(\left(1 + \frac{\alpha}{2}\right)^{I/2} - 1\right)\mathrm{vol}(L_{j^*+\lceil I/2 \rceil + 1}) \geq \left(1 + \frac{\alpha}{2}\right)^{I/2} - 1 \geq \left(1 + \frac{\alpha}{2}\right)^{\frac{2.5}{\alpha}\log(3\mathrm{vol}(A)/\sigma)} - 1$$

$$\overset{\text{Lemma 4.4}}{\geq} \left(1 + \frac{\alpha}{2}\right)^{\frac{2.5}{\alpha}\log(\mathrm{vol}(A \cup B_s))} - 1 > e^{\log(\mathrm{vol}(A \cup B_s))} - 1 = \mathrm{vol}(A \cup B_s) - 1$$

contradicting to the fact that $L_{j^*+1}, \ldots, L_{j^*+\lceil I/2 \rceil} \subseteq A \cup B_s$ from Lemma 4.1.

In sum, we conclude that one of the layered cuts $S_j$ must satisfy $\phi(S_j) < 2\alpha$, finishing the proof of the conductance guarantee.

At last, we note that in both cases (whether the $s$-$t$ flow computed is exact or only approximate), the cut produced satisfies $S \subseteq A \cup B_s$ so $\mathrm{vol}(S) \leq \mathrm{vol}(A)(1 + 1/\varepsilon_\sigma)$ according to Lemma 4.4 again. The running time of $\texttt{LocalFlow}_G(A, \alpha, \varepsilon_\sigma)$ is a direct consequence of

- our choice of $I = \lceil \frac{5}{\alpha}\log(3\mathrm{vol}(A)/\sigma) \rceil$,
- $\mathrm{vol}(A \cup B_s) \leq 3\mathrm{vol}(A)/\sigma$, and
- the running time guarantee from our $\texttt{BlockFlow}$ on the local graph in Lemma 4.3.

$\square$

| Notation | Explanation |
|---|---|
| $\sigma \in (0,1]$ | A lower bound on the size of $\frac{\mathrm{vol}(A \cap S^*)}{\mathrm{vol}(S^*)}$ . |
| $\alpha \in (0,1]$ | A parameter to construct the augmented graph $G' = G_A(\alpha, \varepsilon_\sigma)$. |
| $\varepsilon_\sigma \stackrel{\text{def}}{=} \frac{1}{3(1/\sigma - 1)}$ $\in \left[\frac{\mathrm{vol}(A)}{\mathrm{vol}(V-A)}, 1\right]$ | A parameter to construct the augmented graph $G' = G_A(\alpha, \varepsilon_\sigma)$. |
| $\varepsilon \in (0,1]$ | The stopping rule for binary search in `LocalImprove` and `LocalImprove`$^{\text{exact}}$. |
| $l$ and $\widetilde{l}$ | The length functions used by Goldberg and Rao's algorithm Appendix A.2. |
| $\widehat{l}$ | Our length function Appendix B.1. |
| $\Delta > 0$ | A parameter used by Goldberg and Rao's algorithm to construct length function $l$, and will be halved at each outer iteration; see Algorithm 3. |

Table 1: Summary of Terminology.

## Acknowledgements

We thank Jon Kelner, Silvio Lattanzi, Vahab Mirrokni, Satish Rao and Luca Trevisan for helpful conversations. This material is based upon work partly supported by the National Science Foundation under Grant CCF-1319460 and by a Simons Award (grant no. 284059).

# Appendix

# A    Preliminaries on Network Flow Algorithms

In this section we are interested in the *s-t* maximum flow problem on an $n$-vertex $m$-edge graph $G$ with probably non-unit edge capacities and directed edges.

## A.1    Blocking Flow and Dinic's Algorithm

We first review Dinic's algorithm for computing the maximum flow, and it is based on a concept called *blocking flow* that needs to be computed on the *admissible graph*.

**Residual Graph.**    Given graph $G = (V, E)$ along with a valid *s-t* flow $f$ respecting to the capacities, one can define the *residual graph relative to $f$* as $G_f = (V, E_f)$, where $E_f$ contains all directed edges $(u,v)$ for which $f_{uv} < c_{uv}$. Here $f_{uv}$ is the amount of flow that goes from $u$ to $v$ in $f$, and may be negative if the actual flow goes from $v$ to $u$; $c_{uv}$ is the capacity upper bound in $G$ in the direction from $u$ to $v$ and may be zero. We say that edge $(u,v) \in E_f$ has a *residual capacity* of $c_{uv} - f_{uv} > 0$. An edge $(u,v) \in E$ that is not present in $E_f$ because $f_{uv} = c_{uv}$ is called a *saturated edge*.

**Admissible Graph.**    Given $G$ and a valid *s-t* flow $f$, we let $d(u)$ be the distance of from vertex $s$ to $u$ in the residual graph $G_f = (V, E_f)$, assuming that each edge in the residual graph has length 1. Now a directed edge $(u,v) \in E_f$ always satisfies $d(u) + 1 \geq d(v)$ by the definition of distance labeling $d$, and is called *admissible* if it satisfies $d(u) + 1 = d(v)$. We denote by $G_f^a = (V, E_f^a)$ the admissible graph constructed from $G_f$ but keeping only those admissible edges. It is not hard to see that $G_f^a$ is a layered graph and thus acyclic.

**Blocking Flow.** The blocking flow problem aims to find a (not necessarily maximum) $s$-$t$ flow $f$ on an $n$-vertex $m$-edge acyclic graph, such that in the residual graph $s$ and $t$ are disconnected. In other words, in the original graph every $s$-$t$ path contains at least one saturated edge. Sleator and Tarjan [ST83] proposed a link-cut tree data structure that gives a strongly polynomial $O(m \log n)$ time bound for the blocking flow problem.

**Dinic's Algorithm.** Now we are ready to state Dinic's algorithm [Din70]. It is a simple iterative algorithm that starts with a zero flow $f$. At each iteration, it computes a blocking flow in graph $G_f^a$, denoted by $\texttt{BlockFlow}_{G,f}(s,t)$, and adds it to the current flow $f$. It terminates whenever the blocking flow fails to augment. We state without proof the following two important properties about Dinic's algorithm.

**Proposition A.1.** *In Dinic's algorithm, let $d^{(i)}$ be the distance labeling $d$ for the residual graph after the augmentations of first $i$ blocking flows. Then it satisfies that:*

- *$d^{(i+1)}(u) \geq d^{(i)}(u)$ for each $u \in V$; and*

- *$d^{(i+1)}(t) \geq d^{(i)}(t) + 1$ for the sink $t$, and thus $d^{(i)}(t) \geq i + 1$.*

As a result of the above proposition, Dinic's algorithm must terminate in $n$ steps (because the sink $t$ can be at most at distance $n$ away from the source $s$), and when it terminate $s$ and $t$ are already disconnected in the residual graph $G_f$ so $f$ is already the $s$-$t$ maximum flow. This algorithm can be implemented to run in time $O(mn \log n)$.

## A.2 Goldberg and Rao's Algorithm

In this subsection we briefly review Goldberg and Rao's algorithm [GR98] for computing exact $s$-$t$ flows. (This is a weakly polynomial time algorithm so that a factor of $\log U$ will appear in the running time, where $U$ is an integral upper bound on the maximum flow value and all edge capacities need to be integral.)

It attempts to compute blocking flow iteration by iteration, just like Dinic's. However, unlike Dinic who assigns length 1 to all edges in the residual graph and them computes $d(u)$ as the shortest path distance from $s$ to $u$, Goldberg and Rao choose certain edges in the residual graph and give them length 0 instead.

More specifically, given some parameter $\Delta > 0$ to be chosen later, they defined two binary length functions, $l$ and $\widetilde{l}$. For an edge $(u,v)$ in the residual graph $G'_f$, let

$$l_{uv} = \begin{cases} 0, & \text{if the residual capacity } c_{uv} - f_{uv} \geq 3\Delta \\ 1, & \text{otherwise.} \end{cases}$$

Let $d(u)$ be the shortest path distance in $G_f$ from $s$ to $u$ under this length function $l$. In addition, they define $\widetilde{l}$ on $G_f$ based on $l$. For an edge $(u,v)$ in the residual graph $G_f$

$$\widetilde{l}_{uv} = \begin{cases} 0, & \text{if } d(i) = d(j),\ 2\Delta \leq c_{uv} - f_{uv} < 3\Delta,\ \text{and } c_{vu} - f_{vu} \geq 3\Delta \\ l_{uv}, & \text{otherwise.} \end{cases} \tag{A.1}$$

Directed edges whose lengths are reset to zero in the above formula are called *special* edges. It is a simple exercise to verify that $d$ is also the shortest path distance under length function $\widetilde{l}$.

Under this new length function $\widetilde{l}$, the admissible graph can be defined in the similar way. An edge $(u,v) \in G_f$ satisfying $d(u) + \widetilde{l}(u,v) = d(v)$ is called admissible. Unfortunately, this graph is not necessarily acyclic anymore, so one cannot directly rely on the classical blocking flow algorithm. The key idea from Goldberg and Rao is to provide an $O(|E|\log|V|)$-time algorithm $\mathtt{BinaryBlockFlow}_{G,f,\widetilde{l},\Delta}(s,t)$ that

- either produces an augmenting flow with value $\Delta$, or
- finds a blocking from on this admissible graph.

This algorithm is called the *binary blocking flow* algorithm, because all edge lengths are binary. It is implemented by first shrinking strongly-connected components of the graph by length-0 edges, and then running the classical blocking flow algorithm on the remaining acyclic graph. For the purpose of our paper, we are not interested in the actual implementation of this step. We refer interested readers to Section 5 of [GR98].

Now we are ready to sketch the algorithm of Goldberg and Rao. As described in Algorithm 3, they first define $\Lambda = m^{1/2}$, a term that will appear very often. Initially $F$ is assigned to be the upper bound of the maximum flow value, and $f$ is initialized to an empty flow.

---

**Algorithm 3** Goldberg-Rao's Maximum Flow Algorithm

**Input:** $G = (V,E)$ that includes a source $s$ and a sink $t$.
**Output:** $f$ the maximum $s$-$t$ flow.
1: $\Lambda = m^{1/2}$.
2: $F \leftarrow U$, an upper bound on the maximum flow value.
3: $f \leftarrow 0$.
4: **while** $F \geq 1$ **do**
5:     $\Delta \leftarrow \frac{F}{6\Lambda}$.
6:     **for** $i \leftarrow 1$ to $4\Lambda$ **do**
7:         $\widetilde{l} \leftarrow$ the length function from (A.1).
8:         $f \leftarrow f + \mathtt{BinaryBlockFlow}_{G,f,\widetilde{l},\Delta}(s,t)$.
9:     **end for**
10:    $F \leftarrow F/2$.
11: **end while**
12: **return** $f$.

---

The algorithm will ensure that $F$ is always an upper bound on the difference between the maximum flow value, and the current flow value on $f$. This is true at the beginning. In each outer iteration, we notice that $4\Lambda$ binary blocking flows are computed. There are two possibilities:

- We find a flow of value $\Delta$ at least $3\Lambda$ times.

  In this case we have increased the flow value of $f$ by $\Delta \cdot 3\Lambda = \frac{F}{2}$, so the new difference between the maximum flow value and $f$ is upper bound by $\frac{F}{2}$.

- We find a blocking flow at least $\Lambda$ times.

  In this case the $s$-$t$ distance $d(t)$ increases at least by one per blocking flow augmentation[10], and thus $d(t) \geq \Lambda$. By an averaging argument, there exists a cut in the residual graph of at most $m/\Lambda = \Lambda$ edges. Since each edge of length 1 has a residual capacity of at least $3\Delta$ according to (A.1), this cut carries a total residual capacity of at most $\Lambda \cdot 3\Delta = \frac{F}{2}$ total capacity. It therefore suggests that the maximum flow in the residual graph is upper bounded by $\frac{F}{2}$.

---

[10]This requires a slightly stronger argument than Proposition A.1, and we refer interested readers to [GR98, Lemma 4.1 and Theorem 4.3].

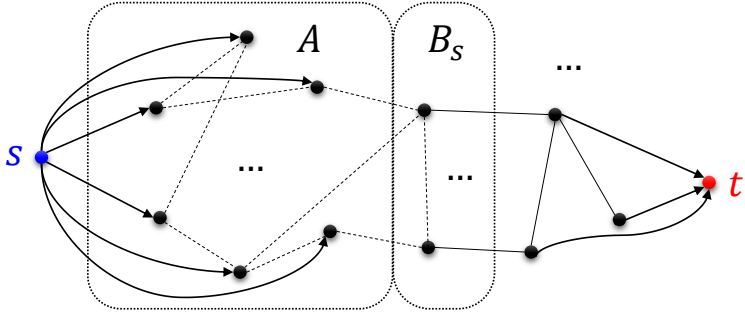

Figure 4: The illustration of modern and classical edges in the residual graph $G'_f$. Dashed edges are modern and solid ones are classical.

In either case, we have the guarantee that difference between the maximum $s$-$t$ flow value and the current flow value $f$ is upper bounded by $F/2$, so we can go to the next outer iteration by halving $F$. The correctness of this algorithm follows from the fact that the outer loop terminates when $F < 1$, and the running time is $O(m^{3/2} \log m \log U)$.

## B  Local Improvement Using Exact Maximum Flow

In this section we prove Theorem 1b.

Recall that in Section 4.2, we have obtained a local but approximate (due to early termination) version of Dinic's algorithm on the augmented graph $G' = G_A(\alpha, \varepsilon_\sigma)$, with a running time $\widetilde{O}\left(\frac{\operatorname{vol}(A)}{\sigma\alpha}\right)$. This running time has an inverse dependency on the target conductance $\alpha$. Can we avoid such dependency by making use of Goldberg and Rao's $\widetilde{O}(m^{1.5})$-time exact maximum-flow algorithm [GR98]? The answer turns out to be affirmative, and we show that, *a variant* of Goldberg and Rao's algorithm yields a local but *exact* maximum flow algorithm on $G' = G_A(\alpha, \varepsilon_\sigma)$, with a running time $\widetilde{O}\left(\left(\frac{\operatorname{vol}(A)}{\sigma}\right)^{1.5}\right)$ not dependent on $\alpha$.

### B.1  Our New Length Function

More precisely, to preserve locality of the running time, we need to study a mixture of Dinic's and Goldberg-Rao's algorithm as follows. For an edge $(u, v)$ in the residual graph $G'_f$ relative to some current $s$-$t$ flow $f$, instead of assigning its length to 1 according to Dinic, or assigning its length to $\widetilde{l}_{uv} \in \{0, 1\}$ according to Goldberg and Rao (for the definition of $\widetilde{l}$ see Appendix A.2), we consider two cases. Letting $B_s \subseteq V - A$ be the set of vertices whose edges to the sink $t$ are fully saturated, for an edge $(u, v) \in G'_f$ in the residual graph (see Figure 4):

- if $u, v \in A \cup B_s$, we say that $(u, v)$ is <u>*modern*</u> and let its length be $\widehat{l}_{uv} = \widetilde{l}_{uv}$; and
- otherwise, we say that $(u, v)$ is <u>*classical*</u> and let its length be $\widehat{l}_{uv} = 1$.

It is important to notice that edges in the residual graphs with length 0 must be modern.

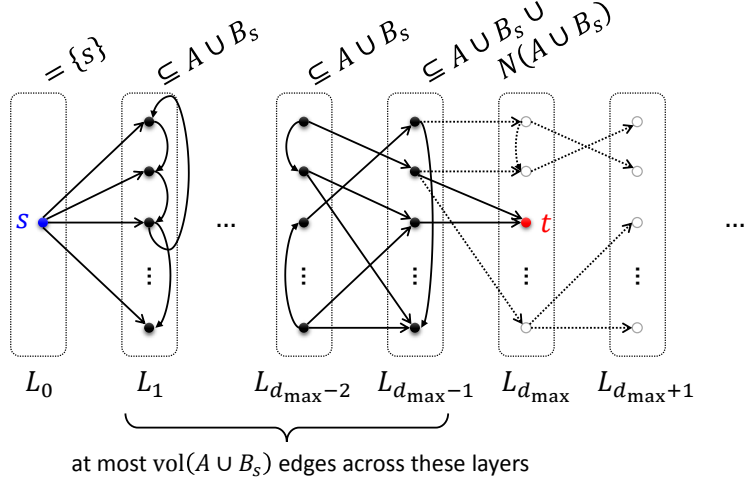

Figure 5: The layers of the (possibly cyclic) admissible graph of Goldberg and Rao's algorithm using our length function $\widehat{l}$. The set inclusions follow from Lemma B.1. The solid edges and vertices are part of the local graph $G'' = G'\langle B_s \rangle$.

Let $d(u)$ be the distance from the source $s$ to $u$ in the residual graph $G'_f$ according to this mixed length function $\widehat{l}$, one can study the admissible graph with only residual edges $(u, v)$ satisfying $d(u) + \widehat{l}_{uv} = d(v)$. Just like in Goldberg and Rao's case, this admissible graph is cyclic but their binary blocking flow algorithm $\texttt{BinaryBlockFlow}_{G', f, \widehat{l}, \Delta}(s, t)$ can find (see Appendix A.2)

- either an $s$-$t$ augmenting flow with value $\Delta$, or
- a blocking from on this admissible graph,

in a running time of $O(m \log n)$. We are now ready to make this binary blocking flow step local.

## B.2 Localizing Binary Blocking Flow with Length Function $\widehat{l}$

Unlike the length function $\widetilde{l}$ used in Goldberg and Rao's, we are able to localize the binary blocking flow algorithm with respect to our new function $\widehat{l}$. Similar to Section 4.1, let us define

- $B_s \subseteq V - A$ to be the set of vertices whose edges to $t$ are already saturated,
- $d(v)$ to be the shortest path distance from $s$ to $v$ in the residual graph of $G'_f$ according to $\widehat{l}$,
- $L_j = \{v \in G' \mid d(v) = j\}$ to be the $j$-th layer of distance $d$, and
- $d_{\max} \overset{\text{def}}{=} d(t)$ to be the distance between the source and the sink.

And we show that the same upper bound Lemma 4.1 holds: (and we emphasize here that this upper bound is not going to hold if $\widetilde{l}$ is used)

**Lemma B.1** (cf. Lemma 4.1). *For every $i \geq 0$, if $d_{\max} < \infty$, we have*

- $d_{\max} \geq 3$.
- $L_j = \{s\}$ *for $j = 0$.*
- $L_j \subseteq A \cup B_s$ *for every $1 \leq j \leq d_{\max} - 2$.*
- $L_j \subseteq A \cup B_s \cup N(A \cup B_s)$ *for $j = d_{\max} - 1$.*

*Proof.* We first show that $d_{\max} \geq 3$ under our definition of $\widehat{l}$. It is clear that $d_{\max} \geq 2$ because all edges from $s$ to $A$ and all edges from $V - A$ to $t$ are *classical*, and thus of length 1. Suppose now that there is a path of length 2 connecting $s$ to $t$: $s \to u_1 \to \ldots u_k \to t$. We automatically have that the lengths between consecutive $u_j$ and $u_{j+1}$ must be zero, and therefore the edges connecting them are *modern*. This indicates that $u_k \in B_s$ is saturated at this step, contradicting to the fact that $u_k \to t$ is an edge in the residual graph $G'_f$.

We now only need to verify the last two items.

For every $1 \leq j \leq d_{\max} - 2$ and an arbitrary $v \in L_j$, we need to show that $v \in A \cup B_s$. If $v \in A$ then we are done, so let us focus on the case when $v \in V - A$ so there is an edge from $v$ to $t$ in $G'$ of capacity $\varepsilon_\sigma \deg(v) > 0$ by the definition of $G'$. In fact, because $d(v) = j \leq d(t) - 2$, it must be true that this edge $(v, t)$ is saturated already, and therefore $v \in B_s$. This proves the third item.

For $j = d_{\max} - 1$ and an arbitrary $v \in L_j$, we need to show that $v \in A \cup B_s \cup N(A \cup B_s)$. If $v \in A \cup B_s$ then we are done. Otherwise, there must exist an edge $(u, v) \in G'_f$ in the residual graph such that $d(u) + \widehat{l}_{uv} = d(v)$ where $\widehat{l}$ is the length function at this iteration. Since $v \notin A \cup B_s$, this residual edge $(u, v)$ must be *classical*, and therefore $\widehat{l}_{uv} = 1$ so vertex $u \in L_{j-1}$ where $j - 1 \geq 1$. By the previous case, we have that $u \in A \cup B_s$. This concludes that $v$ is a neighbor to some vertex in $A \cup B_s$, and therefore $v \in A \cup B_s \cup N(A \cup B_s)$, proving the last item. $\square$

Notice that the proof above does not require any property on $\Delta$, the parameter that was used to define the length function $\widetilde{l}$, and therefore also $\widehat{l}$.

We next show that similar to Lemma 4.3, the binary blocking flow can also be computed locally.

**Lemma B.2** (cf. Lemma 4.3). *When computing* $\mathtt{BinaryBlockFlow}_{G', f, \widehat{l}, \Delta}(s, t)$, *it suffices to compute* $\mathtt{BinaryBlockFlow}_{G'', f, \widehat{l}, \Delta}(s, t)$ *on the local graph* $G'' = G'\langle B_s \rangle$ *(recall Definition 4.2). Therefore, the running time is* $O\big(\mathrm{vol}(A \cup B_s) \log \mathrm{vol}(A \cup B_s)\big)$.

*Proof.* The first half of the proof of Lemma 4.3 still applies: $s$ and $t$ are disconnected in the residual graph $G'_f$ if and only if they are disconnected in $G''_f$. Therefore, for the rest of the proof, we can focus on $d(t) = d_{\max} < \infty$.

Recall that $L_0, L_1, \ldots, L_{d_{\max}}$ are the layers of vertices in the residual graph $G'_f$ that have distance $0, 1, \ldots, d_{\max}$ respectively from the source $s$. Since the sink $t$ has a distance $d(t) = d_{\max}$, there is no need for the binary blocking flow to ever touch any vertex $v$ with $d(v) \geq d_{\max}$ except $t$ itself. In other words, the vertices in the support of the binary blocking flow must be a subset of

$$\{s, t\} \cup L_1 \cup \cdots \cup L_{d_{\max} - 1} \subseteq \{s, t\} \cup A \cup B_s \cup N(A \cup B_s) \ ,$$

(see the solid vertices in Figure 5 and recall Lemma B.1) and are thus in $G''$.

We then turn to the edges in the support of $\mathtt{BinaryBlockFlow}_{G', f, \widehat{l}, \Delta}(s, t)$. By the definition of blocking flow, those edges must be (see the solid edges in Figure 5 and recall Lemma B.1):

- either between layers
  - from $s$ to $L_1$, which is of the form $\{(s, v) : v \in A\}$;
  - from $L_{d_{\max} - 1}$ to $t$, which is of the form $\big\{(v, t) : v \in B_s \cup N(A \cup B_s)\big\}$;
  - from $L_j$ to $L_{j+1}$ for $1 \leq j \leq d_{\max} - 2$, which is of the form $\big\{(u, v) : u \in A \cup B_s, v \in V\big\}$;
- or within layers from $L_j$ to $L_j$, so those edges have length 0 and must be *modern* according to the definition of $\widehat{l}$, and thus of the form $\big\{(u, v) : u \in A \cup B_s, v \in A \cup B_s\big\}$.

In sum, the edges in the support of $\texttt{BinaryBlockFlow}_{G',f,\widehat{l},\Delta}(s,t)$ are also in $G''$. This essentially ends the proof because when computing $\texttt{BinaryBlockFlow}_{G',f,\widehat{l},\Delta}(s,t)$, it suffices to compute $\texttt{BinaryBlockFlow}_{G'',f,\widehat{l},\Delta}(s,t)$ on $G''$ which guarantees to contain the support of the blocking flow.

At last we focus on the running time. The total number of edges in $G''$ is $O\big(\text{vol}(A \cup B_s)\big)$, and therefore the running time for $\texttt{BinaryBlockFlow}_{G'',f,\widehat{l},\Delta}(s,t)$ is $O\big(\text{vol}(A \cup B_s) \log \text{vol}(A \cup B_s)\big)$ owing to Appendix A.2. □

## B.3 Localizing Goldberg-Rao's Algorithm

In this section we want to localize Goldberg-Rao's algorithm to compute the exact $s$-$t$ maximum flow on our augmented graph $G' = G_A(\alpha, \varepsilon_\sigma)$.

Recall from Appendix A.2 that Goldberg and Rao's algorithm is nothing but a sequence of binary blocking flow computations, where the length function varies between each consecutive runs. Also, after every $4\Lambda$ such computations, the value of $\Delta$ will be halved. We now rewrite this algorithm, using our new length function $\widehat{l}$ defined in Appendix B.1, and our localized binary blocking flow algorithm in Appendix B.2. We summarize our algorithm as $\texttt{LocalFlow}^{\texttt{exact}}$ in Algorithm 4, and prove the following lemma on its running time

**Lemma B.3.** Let $(f, S) = \texttt{LocalFlow}_G^{\texttt{exact}}(A, \alpha, \varepsilon_\sigma)$, then $f$ and $S$ are the $s$-$t$ maximum flow and minimum cut respectively in $G' = G_A(\alpha, \varepsilon_\sigma)$. In addition, it runs in time

$$O\left(\frac{\text{vol}(A)^{1.5}}{\sigma^{1.5}} \log^2\left(\frac{\text{vol}(A)}{\sigma}\right)\right) \quad .$$

The rest of this subsection is devoted to proving this lemma.

---

**Algorithm 4** $\texttt{LocalFlow}_G^{\texttt{exact}}(A, \alpha, \varepsilon_\sigma)$

**Input:** $G = (V, E)$, $A \subset V$, $\alpha \in (0, 1]$, and $\varepsilon_\sigma \in \left[\frac{\text{vol}(A)}{\text{vol}(V-A)}, \infty\right)$.
**Output:** the $s$-$t$ maximum flow $f$ and its dual $s$-$t$ minimum cut $S$ in $G_A(\alpha, \varepsilon_\sigma)$.
1: $G' \leftarrow G_A(\alpha, \varepsilon_\sigma)$.                                          {see Definition 3.1.}
2: $\Lambda = (3\text{vol}(A)/\sigma)^{1/2}$.
3: Scale $G'$ up so that all capacities are integral, and $F \leftarrow \texttt{poly}(\frac{\text{vol}(A)}{\sigma})$ is an upper bound on the $s$-$t$ flow value.
4: $B_s \leftarrow \varnothing$.
5: $f \leftarrow 0$.
6: **while** $F \geq 1$ **do**
7:     $\Delta \leftarrow \frac{F}{4\Lambda}$.
8:     **for** $i \leftarrow 1$ **to** $6\Lambda$ **do**
9:         $\widehat{l} \leftarrow$ the length function from Appendix B.1.
10:         $G'' \leftarrow G'\langle B_s \rangle$.                                   {see Definition 4.2.}
11:         $f \leftarrow f + \texttt{BinaryBlockFlow}_{G'',f,\widehat{l},\Delta}(s,t)$.
12:         $C \leftarrow$ the vertices in $N(A \cup B_s)$ whose edges to the sink get saturated in the new flow
13:         $B_s \leftarrow B_s \cup C$.
14:     **end for**
15:     $F \leftarrow F/2$.
16: **end while**
17: **return** the maximum $s$-$t$ flow $f$ and its corresponding minimum cut $S$ (after scaling down).

---

**Integrality.** For our application of computing the exact $s$-$t$ maximum flow, recall that although we may have edges with fractional capacities, all of our capacities values in $G'$ are either a multiple of $\alpha$ or $\varepsilon_\sigma$, which both can be made into rational numbers with polynomial-sized numerators and denominators (in terms of $\frac{\mathrm{vol}(A)}{\sigma}$), so with simple scaling we can assume that our graph $G'$ is of integral capacities and has a maximum flow upper bound $\mathsf{poly}(\frac{\mathrm{vol}(A)}{\sigma})$. This will introduce an extra factor of $\log(\frac{\mathrm{vol}(A)}{\sigma})$ in the running time since Goldberg and Rao's algorithm is weakly polynomial.

**Correctness.** We now argue that our $\mathtt{LocalFlow}^{\mathsf{exact}}$ performs as well as the original Goldberg and Rao's, even though our length function is changed from $\widetilde{l}$ to $\widehat{l}$. Again, we will ensure that $F$ is always an upper bound on the maximum possible flow one can push in the residual graph $G'_f$.

Indeed, for each outer iteration, we notice $F$ and $\Delta$ are fixed and $4\Lambda$ binary blocking flows are computed. There are two possibilities:

- We find a flow of value $\Delta$ at least $3\Lambda$ times, in which case we have increased the flow value by $\Delta \cdot 3\Lambda = \frac{F}{2}$. In this case we can successfully go to the next outer iteration with $F \leftarrow F/2$.

- We find a blocking flow at least $\Lambda$ times.

In this later case, we make use of the following lemma which can be similarly proved just like the classical Goldberg and Rao's:

**Lemma B.4.** *Let $d$ and $\widehat{l}$ be the distance labeling and length function before a binary blocking flow computation, and $d'$ and $\widehat{l}'$ be the distance labeling and length function immediately after it. Then:*

- *$d$ is a distance labeling with respect to $\widehat{l}'$, so $d(u) \leq d'(u)$ for all $u$.*

- *If the binary blocking flow computation ends up producing a blocking flow, then $d(t) < d'(t)$.*

*Proof.* Ignored in the current version because it is a simple repetition of Lemma 4.1 and Theorem 4.3 in [GR98]. $\square$

This lemma indicates that, the $s$-$t$ distance $d(t)$ increases at least by one per blocking flow augmentation, and does not decrease if $\mathtt{BinaryBlockFlow}$ only finds a flow of value $\Delta$. Since we have $d(t) \geq 3$ at the beginning of this $4\Lambda$ iterations (recall Lemma B.1), this shows that $d(t) \geq \Lambda + 3$ at the end of the execution because $\Lambda$ blocking flows are found.

Now let us denote by $d$ the this distance labeling after the execution of an outer layer (i.e., $4\Lambda$ binary blocking flow computations), and let $L_0, L_1, L_2, \ldots, L_{d_{\max}}$ be the layers of vertices with distance $0, 1, 2, \ldots, d_{\max} = d(t) \geq \Lambda + 3$.

Now we only focus on the cuts between layers $L_1, L_2, \ldots, L_{d_{\max}-2}$. It follows from Lemma B.1 that the edges between consecutive layers are inside $E(A \cup B_s, A \cup B_s)$, so there are at most $\mathrm{vol}(A \cup B_s) \leq \frac{3}{\sigma}\mathrm{vol}(A)$ of them, according to Lemma 4.4.

If $\Lambda = (3\mathrm{vol}(A)/\sigma)^{1/2}$, then among those $d_{\max} - 3 \geq \Lambda$ layer cuts, by an averaging argument there exist one, say $L_j$ and $L_{j+1}$, satisfies that the total number of residual edges between $L_j$ and $L_{j+1}$ is upper bounded by $\frac{\mathrm{vol}(A \cup B_s)}{\Lambda} = \Lambda$. They carry a total residual capacity of $\Lambda\Delta = \frac{F}{2}$ because these edges are modern and modern edges with length 1 must have residual capacity at most $3\Delta$ (see (A.1)), this gives a total residual capacity of $\Lambda \cdot 3\Delta = \frac{F}{2}$. Therefore, we can arrive at the same conclusion as Goldberg and Rao: the remaining flow in the residual graph is at most $\frac{F}{2}$ so it is an upper bound between the current and maximum $s$-$t$ flow values in $G'_f$.

This essential ends the proof of correctness because at the end of the iteration $F < 1$.

**Running Time.** Since each binary blocking flow with our special length function $\widehat{l}$ can be computed in time $O\big(\mathrm{vol}(A \cup B_s)\log \mathrm{vol}(A \cup B_s)\big) = O\left(\frac{\mathrm{vol}(A)}{\sigma}\log\left(\frac{\mathrm{vol}(A)}{\sigma}\right)\right)$, and we have a total of $O(\log\frac{\mathrm{vol}(A)}{\sigma}\cdot\Lambda)$ such computations, this gives a total running time of:

$$O\left(\frac{\mathrm{vol}(A)^{1.5}}{\sigma^{1.5}}\log^2\left(\frac{\mathrm{vol}(A)}{\sigma}\right)\right) \ .$$

This ends the proof of Lemma B.3. Since we are now able to compute the exact *s-t* max-flow-min-cut value in the augmented graph $G' = G_A(\alpha, \varepsilon_\sigma)$ locally, we can replace Lemma 4.5 by the following (slightly stronger) lemma whose proof is even simpler and is a direct consequence of Lemma 3.2 and Lemma B.3.

**Lemma B.5** (`LocalFlow`$^{\mathsf{exact}}$ performance). *Given arbitrary $\alpha \in (0, 1]$, $\varepsilon_\sigma \in [\frac{\mathrm{vol}(A)}{\mathrm{vol}(V-A)}, \infty)$, and suppose there is some set $S^*$ satisfying:*

$$\frac{|E(S^*, V - S^*)|}{\mathrm{vol}(S^*)} < \alpha\left(\frac{\mathrm{vol}(A \cap S^*)}{\mathrm{vol}(S^*)} - \varepsilon_\sigma\frac{\mathrm{vol}(S^* - A)}{\mathrm{vol}(S^*)}\right) \ . \quad\quad\quad \text{(same as (4.1))}$$

*Then,* `LocalFlow`$_G^{\mathsf{exact}}(A, \alpha, \varepsilon_\sigma)$ *outputs a flow $f$ with value strictly smaller than $\mathrm{vol}(A)$, and produces a set $S$ of size $\mathrm{vol}(S) \le \mathrm{vol}(A)(1 + 1/\varepsilon_\sigma)$ with conductance*

$$\phi(S) = \frac{|E(S, V - S)|}{\min\{\mathrm{vol}(S), \mathrm{vol}(V - S)\}} < \alpha \ .$$

*The running time of* `LocalFlow`$_G^{\mathsf{exact}}(A, \alpha, \varepsilon_\sigma)$ *is $O\left(\frac{\mathrm{vol}(A)^{1.5}}{\sigma^{1.5}}\log^2\left(\frac{\mathrm{vol}(A)}{\sigma}\right)\right)$.*

## B.4 The Final Algorithm

Now we are ready to put everything together and construct our final algorithm for local cut improvement using our proposed `LocalFlow`$^{\mathsf{exact}}$ algorithm. If one replaces the use of Lemma 4.5 by Lemma B.5 in the proof of Theorem 4.6, he can immediately obtain the following

**Theorem B.6.** *Given arbitrary $\varepsilon_\sigma \in [\frac{\mathrm{vol}(A)}{\mathrm{vol}(V-A)}, \infty)$ and $\varepsilon \in (0, 1]$, and suppose there is some set $S^*$ satisfying:*

$$\frac{|E(S^*, V - S^*)|}{\mathrm{vol}(S^*)} < \alpha^*\left(\frac{\mathrm{vol}(A \cap S^*)}{\mathrm{vol}(S^*)} - \varepsilon_\sigma\frac{\mathrm{vol}(S^* - A)}{\mathrm{vol}(S^*)}\right) \ . \quad\quad\quad \text{(same as (4.1))}$$

*Then,* `LocalImprove`$_G^{\mathsf{exact}}(A, \varepsilon_\sigma, \varepsilon)$ *outputs a set $S$ of size $\mathrm{vol}(S) \le \mathrm{vol}(A)(1 + 1/\varepsilon_\sigma)$ with conductance*

$$\phi(S) = \frac{|E(S, V - S)|}{\min\{\mathrm{vol}(S), \mathrm{vol}(V - S)\}} < (1 + \varepsilon)\alpha^* \ .$$

*In addition, the running time is*

$$O\left(\frac{\mathrm{vol}(A)^{1.5}}{\sigma^{1.5}}\log^2\left(\frac{\mathrm{vol}(A)}{\sigma}\right)\log\left(\frac{1}{\varepsilon}\right)\right) = \widetilde{O}\left(\frac{\mathrm{vol}(A)^{1.5}}{\sigma^{1.5}}\right) \ .$$

Now our Theorem 1b is a direct corollary of the above (and in fact, more general) theorem.

---

**Algorithm 5** `LocalImprove`$_G^{\text{exact}}(A, \varepsilon_\sigma, \varepsilon)$

---

**Input:** $G = (V, E)$, $A \subset V$, $\varepsilon_\sigma \in \left[\frac{\text{vol}(A)}{\text{vol}(V-A)}, \infty\right)$, $\varepsilon \in (0, 1]$.

**Output:** a non-empty set $S \subset V$ with good cut conductance.

 1: $\alpha_{\min} \leftarrow 0, \alpha_{\max} \leftarrow 1$.
 2: **while** $\alpha_{\max} - \alpha_{\min} > \varepsilon \alpha_{\min}$ **do**
 3:     $\alpha \leftarrow \frac{1}{2}(\alpha_{\min} + \alpha_{\max})$.
 4:     **if** the flow value `LocalFlow`$_G^{\text{exact}}(A, \alpha, \varepsilon_\sigma)$ is larger than or equal to $\text{vol}(A)$ **then**
 5:        $\alpha_{\min} = \alpha$.
 6:     **else**
 7:        $\alpha_{\max} = \alpha$.
 8:     **end if**
 9: **end while**
10: $(f, S) \leftarrow$ `LocalFlow`$_G^{\text{exact}}(A, \alpha_{\max}, \varepsilon_\sigma)$.
11: **return** $S$.

---

**Theorem 1b** (restated). *Given set $A \subset V$ of the graph and a constant $\sigma \in (0, 1]$ satisfying $\frac{\text{vol}(V-A)}{\text{vol}(A)} \geq 3(1/\sigma - 1)$, and suppose*

$$\exists \text{ (unknown) target set } S^* \subseteq V \text{ satisfying } \text{vol}(S^*) \leq \text{vol}(V - S^*) \text{ and } \frac{\text{vol}(A \cap S^*)}{\text{vol}(S^*)} = \delta \geq \sigma,$$

*then our* `LocalImprove`$^{\text{exact}}$ *algorithm can be parameterized by $\sigma$ and outputs a set $S$ satisfying*

$$\text{vol}(S) \leq \frac{3}{\sigma}\text{vol}(A) \quad \text{and} \quad \phi(S) \leq \frac{2}{\delta}\phi(S^*)$$

*in time*

$$O\left(\left(\frac{\text{vol}(A)}{\sigma}\right)^{1.5} \log^2\left(\frac{\text{vol}(A)}{\sigma}\right)\right) \ .$$

*Proof.* Recall that we have chosen $\varepsilon_\sigma = \frac{1}{3(1/\sigma - 1)} \geq \frac{\text{vol}(A)}{\text{vol}(V-A)}$, and now let $\alpha = \phi(S^*) \cdot 5/3\sigma$ we have

$$\frac{|E(S^*, V - S^*)|}{\text{vol}(S^*)} = \phi(S^*) = \frac{3\alpha\sigma}{5} < \frac{2\alpha\sigma}{3} \leq \frac{2\alpha}{3}\frac{\text{vol}(A \cap S^*)}{\text{vol}(S^*)}$$

$$= \alpha\left(\frac{\text{vol}(A \cap S^*)}{\text{vol}(S^*)} - \frac{1}{3}\frac{\text{vol}(A \cap S^*)}{\text{vol}(S^*)}\right)$$

$$\leq \alpha\left(\frac{\text{vol}(A \cap S^*)}{\text{vol}(S^*)} - \varepsilon_\sigma(1/\sigma - 1)\frac{\text{vol}(A \cap S^*)}{\text{vol}(S^*)}\right)$$

$$\leq \alpha\left(\frac{\text{vol}(A \cap S^*)}{\text{vol}(S^*)} - \varepsilon_\sigma\frac{\text{vol}(S^*) - \text{vol}(A \cap S^*)}{\text{vol}(S^*)}\right)$$

$$= \alpha\left(\frac{\text{vol}(A \cap S^*)}{\text{vol}(S^*)} - \varepsilon_\sigma\frac{\text{vol}(S^* - A)}{\text{vol}(S^*)}\right) \ .$$

This satisfies the requirement of Theorem B.6, and letting $\varepsilon = 1/5$, `LocalImprove`$^{\text{exact}}$ guarantees to output a set $S$ with conductance $\phi(S) \leq (1 + 1/5)\alpha = \frac{2}{\sigma}\phi(S^*)$, and size $\text{vol}(S) \leq \text{vol}(A)(1 + 1/\varepsilon_\sigma) \leq \frac{3}{\sigma}$. The running time follows by substituting our choices of $\varepsilon_\sigma, \varepsilon$ into Theorem B.6. $\qquad\square$

# C   Certifying Expansion using Local Flows

In this section we restate our core lemmas Lemma 4.5 and Lemma B.5 in their stronger forms using flow certificates.

**Definition C.1.** *Given vertex subset $A \subset V$ and constants $c_1 > 0$ and $c_2 \geq \frac{c_1 \text{vol}(A)}{\text{vol}(B)}$, we define a bipartite demand of $\mathsf{BiDemand}(A, c_1, c_2)$ to be the one that requires to route flows from $A$ to $V - A$: $c_1 \deg(u)$ units of flow need to be routed out of each $u \in A$, and each vertex $v \in V - A$ can receive at most $c_2 \deg(u)$ units of flow.*

It is not hard to verify that if one can route the demand graph $\mathsf{BiDemand}(A, 1, \varepsilon_\sigma)$ in $G$ with congestion $1/\alpha$ using flow $f$, then for all non-empty subset $S \subseteq V$ the flow $f$ certifies a lower bound $|E(S, V - S)| \geq \alpha f(S, V - S)$ where $f(S, V - S)$ is the total amount of flow in $f$ that has source in $S$ and sink in $V - S$. This flow certification view provides sometimes a better lower bound on the conductance of a set, and in the worse case it matches the guarantee in Lemma 3.2 owing to $f(S, V - S) \geq \text{vol}(A \cap S) - \varepsilon_\sigma \text{vol}(S - A)$.

Now we restate Lemma B.5 in this flow certificate view.

**Lemma B.5** (restated). *Given arbitrary $\alpha \in (0, 1]$, $\varepsilon_\sigma \in [\frac{\text{vol}(A)}{\text{vol}(V - A)}, \infty)$, suppose that*

$$\text{all set } S \text{ of size } \text{vol}(S) \leq \text{vol}(A)(1 + 1/\varepsilon_\sigma) \text{ have conductance } \phi(S) \geq \alpha,$$

*then $\mathsf{LocalFlow}_G^{\mathsf{exact}}(A, \alpha, \varepsilon_\sigma)$ produces a flow $f$ that routes the demand graph $\mathsf{BiDemand}(A, 1, \varepsilon_\sigma)$ in $G$ with congestion $1/\alpha$ and runs in time $O\left(\frac{\text{vol}(A)^{1.5}}{\sigma^{1.5}} \log^2 \left(\frac{\text{vol}(A)}{\sigma}\right)\right)$.*

It is interesting to see that Lemma 4.5 can be stated in this flow certificate view with an additional guarantee on the flow path lengths. This type of bounded-length flow certificate may be of independent interest.

**Lemma 4.5** (restated). *Given arbitrary $\alpha \in (0, 1]$, $\varepsilon_\sigma \in [\frac{\text{vol}(A)}{\text{vol}(V - A)}, \infty)$, and suppose that*

$$\text{all set } S \text{ of size } \text{vol}(S) \leq \text{vol}(A)(1 + 1/\varepsilon_\sigma) \text{ have conductance } \phi(S) \geq 2\alpha,$$

*then $\mathsf{LocalFlow}_G(A, \alpha, \varepsilon_\sigma)$ produces a flow $f$ that routes the demand graph $\mathsf{BiDemand}(A, 1, \varepsilon_\sigma)$ in $G$ with congestion $1/\alpha$ and path-length $O\left(\frac{\log(\text{vol}(A)/\sigma)}{\alpha}\right)$, and runs in time $O\left(\frac{\text{vol}(A)}{\sigma\alpha} \log^2 \left(\frac{\text{vol}(A)}{\sigma}\right)\right)$.*